\documentclass[preprinttwo, twocolumn]{aastex63}
\usepackage{amsmath,amssymb,lmodern,mathrsfs,ulem,mathtools,bm}
\usepackage{graphicx}
\usepackage{enumitem}
\usepackage{natbib}
\usepackage{xcolor}
\usepackage{makecell}
\usepackage{hyperref}
\usepackage[modulo]{lineno}
\usepackage[caption=false]{subfig}

\newcommand{\rev}[1]{{#1}}

\definecolor{fgreen}{RGB}{34,139,34}

\journalinfo{Submitted to The Astrophysical Journal 8 April, 2022}
\begin{document}

\revised{16 May, 2022}

\title{The Dynamic Evolution of Solar Wind Streams Following Interchange Reconnection}

\author[0000-0001-8517-4920]{Roger B. Scott}
\affiliation{National Research Council Postdoctoral Fellow at the US Naval Research Laboratory, Washington, DC 20375, USA}

\author[0000-0002-3300-6041]{Stephen J. Bradshaw}
\affiliation{Rice University, Houston, TX 77005, USA}

\author[0000-0002-4459-7510]{Mark G. Linton}
\affiliation{US Naval Research Laboratory, Washington, DC 20375, USA}


\begin{abstract}
\noindent Interchange reconnection is thought to play an important role in determining the dynamics and material composition of the slow solar wind that originates from near coronal hole boundaries.
To explore the implications of this process we simulate the dynamic evolution of a solar wind stream along a newly-opened magnetic flux tube. 
The initial condition is composed of a piecewise continuous dynamic equilibrium in which the regions above and below the reconnection site are extracted from steady-state solutions along open and closed field lines. 
The initial discontinuity at the reconnection site is highly unstable and  evolves as a Riemann problem, decomposing into an outward-propagating shock and inward-propagating rarefaction that eventually develop into a classic N-wave configuration.
This configuration ultimately propagates into the heliosphere as a coherent structure and the entire system eventually settles to a quasi-steady wind solution. 
In addition to simulating the fluid evolution we also calculate the time-dependent non-equilibrium ionization of oxygen in real time in order to construct in situ diagnostics of the conditions near the reconnection site. 
This idealized description of the plasma dynamics along a newly-opened magnetic field line provides a baseline for predicting and interpreting the implications of interchange reconnection for the slow solar wind.
Notably, the density and velocity within the expanding N-wave are generally enhanced over the ambient wind, as is the $\rm O^{7+}/O^{6+}$ ionization ratio, which exhibits a discontinuity across the reconnection site that is transported by the flow and arrives later than the propagating N-wave.\\
\end{abstract}

\section{Introduction}\label{introduction.sec}

One of the more enduring challenges in solar and heliospheric science is to determine what physical processes are responsible for the disparate properties of various solar wind streams.
Compared to the fast solar wind (FSW), the slow solar wind (SSW) exhibits lower typical wind speeds with increased variability \citep{McComas:2000.105}.
Various authors have explored the effects of different types of heating profiles and magnetic geometries in field-aligned (1D) models \citep[e.g.,][and others]{Cranmer:2007, Grappin:2011}, which can produce a wide range of predicted wind speeds but do not predict rapid fluctuations within a single wind stream.
Additionally, the material composition of the SSW shows similarities to the plasma found in active regions, with ionization ratios and elemental abundances that are inconsistent with the coronal-hole regions from which the FSW emanates \citep{Zhao:2017, Laming:2019}.
This suggests that changes in magnetic topology, which are difficult to capture in field-aligned models, may be critical to the properties of the SSW.

The solar corona is partitioned into various spatial domains that reflect the connectivity of the magnetic field.
In regions where the magnetic field maps from the photosphere into the heliosphere and beyond, the associated volume is said to be ``open''.
Conversely, where the field maps between positive and negative polarity domains on the photosphere, the associated volume is said to be ``closed''.
The interfaces between these domains collectively form the open-closed boundary, which is strongly correlated with coronal-hole boundaries in X-ray and EUV images \citep[e.g.,][]{Nikolic:2019}.
According to Alfv\'en's theorem, an ideal plasma (zero electrical resistivity) is coupled to the magnetic field in such a way that individual fluid elements are always connected to the same field line.
Therefore, when the connectivity of the magnetic field is static in time the plasma within the various closed magnetic domains remains confined while the plasma within open domains can stream freely into the heliosphere.

For plasma that originates from within a closed domain to escape across the open-closed boundary into an adjacent coronal hole depends on the process of interchange reconnection \citep{Crooker:2002}, whereby field lines from either side of the open-closed boundary undergo a change in connectivity so that a new flux tube is formed that extends from a footpoint within a previously closed region on the solar surface out into the open coronal hole region and into the heliosphere and beyond.
This process has been studied extensively in 3D magnetohydrodynamic (MHD) simulations of coronal streamers and pseudostreamers \citep[e.g.,][]{Masson:2014j, Higginson:2017a, Aslanyan:2021a, Scott:2021a}; however, these and similar studies have invoked simplistic fluid approximations for computational efficiency.
Therefore, while the magnetic evolution is reasonably well understood, many questions remain concerning the associated plasma dynamics.
\rev{Additionally, while the ionization ratios of trace elements have been studied in detail for steady-state wind streams \citep[see, e.g., ][ and others]{Landi:2012j, Gilly:2020o} little is known about how these evolve during interchange reconnection.
These ratios are often taken as diagnostics of the temperature history from the solar surface up to a given height \citep[the so-called ``freeze-in height'', ][]{Owocki:1983} beyond which the ionization ratios undergo no additional evolution; however, if plasma is released from closed magnetic domains at heights that are comparable to the freeze-in height this could have significant implications for in-situ diagnostics of the solar wind.}

The primary difficulty in constructing an accurate model of interchange reconnection in the context of the solar wind stems from the many decades of spatial and temporal scales that must be simultaneously resolved.
At the largest scale the global magnetic field and location of the acoustic and Alfv\'enic critical points require a numerical domain that covers several tens of $\rm Mm^2$ on the solar surface (i.e., the size of an active region) and extends outward to a height of $20 R_\odot$ or more, meaning that the time required for fluid features to transit the numerical domain is typically on the order of tens of hours for the fastest propagating signals (e.g., electron thermal speed and fast magneto-acoustic mode) and significantly longer for slower modes (e.g., slow magneto-acoustic mode and bulk flow). 
At the opposite extreme, the temperature gradients at the base of the transition region -- which are required to accurately capture the subtle balance of coronal heating, thermal conduction, and radiative losses that dictate the mass flux into the corona -- require a minimum resolution on the order of a few tens of $\rm km$, resulting in a numerical timestep that is typically on the order of $10^{-2} \rm s$.
As a result, constructing a 3D numerical model of interchange reconnection in the presence of a self-consistent solar wind has proven to be computationally prohibitive, and none are currently in common use.

Here we present a hybrid model that partially mitigates the computational demands of a full 3D simulation by computing the field-aligned (1D) plasma dynamics -- including detailed physical processes that structure the transition region -- with an empirical mechanism for emulating the effects of magnetic reconnection. 
Beginning with steady-state solutions for the open- and closed-field regions, we construct an initial condition for the plasma along a newly-opened flux tube such that the fluid properties are discontinuous across the reconnection site, being composed of a transonic wind stream above a hotter and more dense hydrostatic column.
This technique builds on previous work by \cite{Bradshaw:2011d}, but has been extended to include critical physics for the appropriate treatment of an out-flowing wind solution, including modification of the distant outer boundary and variability of the flux-tube cross-sectional area.
By following the hydrodynamic evolution of this newly-formed, post-reconnection plasma column, we are able explore how magnetic reconnection directly affects the plasma evolution when material from disparate source regions are brought into close proximity as the magnetic connectivity changes.

In the following section we describe our simulation design and numerical model. 
Then in section \ref{results.sec} we describe the evolution of the post-reconnection plasma and associated time-dependent ionization of \rev{oxygen}, which we calculate for comparison to in situ studies by \cite{Zhao:2017}.
In section \ref{discussion.sec} we discuss these findings and their implications for in situ measurements as well as the appropriate interpretation of this model in the context of a volume filling, 3D magnetic field. 
Finally, we offer concluding remarks in section \ref{conclusions.sec}.

\section{Simulation Design}\label{design.sec}

\subsection{Fluid Model} \label{sec.fluid}

We model the plasma as a fully-ionized two-temperature fluid with mass density $\rho$, velocity $u$, and individual ion and electron pressures $p_{i}$ and $p_{e}$, and we solve the field-aligned hydrodynamic equations in time ($t$) and distance along the field line ($s$).
The magnetic field geometry is taken to be static in time so that the cross-field dynamics can be ignored.
In the limit that the electron mass is negligible compared to the average ion mass ($m_i \gg m_e$), and assuming quasi-neutrality ($n_e = n_i = n$), the continuity and momentum equations are
\begin{equation}
    \partial_t \rho + \frac{1}{A}\partial_s \left( \rho u A \right) = 0\\
    \label{continuity.eq}
\end{equation} 
and
\begin{equation}
    \partial_t \left(\rho u \right) + \frac{1}{A}\partial_s \left(\rho u^2 A \right) =  \rho g - \partial_s P + \frac{1}{A}\partial_s \left(\sigma A \right),
    \label{momentum.eq}
\end{equation}
where $g$ and $A$ are the gravitational acceleration and cross-sectional area parameterized along $s$, $P = p_{i} + p_{e}$ is the total pressure, and $\sigma$ is the viscous stress.
The conservative form of the energy equations for the ions and electrons are likewise given by
\begin{linenomath}\begin{align}
\partial_t E_{i} & = - \frac{1}{A}
    \partial_s \Big ( (E_{i} + p_{i}) u A + \sigma u A - A F_{i} \Big ) \notag \\
& + \rho u g  - u  \partial_s p_{e} + \frac{k_B n}{\gamma - 1} \nu^{ei} \left (T_{e} - T_{i} \right ) + h_{i}
\label{ion_energy.eq}
\end{align}\end{linenomath}
and 
\begin{linenomath}\begin{align}
\partial_t E_e & = - \frac{1}{A}
    \partial_s \Big ( (E_e + p_e) u A - A F_e \Big ) \notag \\
& + u \partial_s p_e + \frac{k_B n}{\gamma - 1} \nu^{ei} \left (T_i - T_e \right ) - R + h_e,
\label{electron_energy.eq}
\end{align}\end{linenomath}
where the energies are defined as $E_i = \frac{1}{2} \rho u^2 + \frac{1}{\gamma - 1} p_i$ for the ions and $E_e = \frac{1}{\gamma -1} p_e$ for the electrons; $T_i$ and $T_e$ are the ion and electron temperatures; $F_i$ and $F_e$ are the associated conductive heat fluxes; and $h_{i,e}$ and $R$ are the imposed coronal heating and empirical radiative losses, respectively.

We model the radiative losses with a piecewise polynomial fit as described in \citet{Bradshaw:2011d} and we prescribe these to fall off to zero below a minimum temperature of $2\times10^4 {\rm K}$, emulating the effect of an optically thick chromosphere.
The system is closed through the ideal gas law, $p_{i,e} = k_B n T_{i,e}$, and by the definition of the conductive fluxes and ion viscous stress through the Spitzer-Harm formulation; $F_{i,e} = - \kappa_{i,e} T_{i,e}^{5/2} \partial_s T_{i,e}$ and $\sigma = (4/3) \mu_{i} \partial_s u$. 
\rev{The dynamic viscosity is given by $\mu_{i} = \mu_{i}^{(0)} T^{5/2} / \ln \Lambda^{ii}_c$ and ion-electron collision frequency is defined to be $\nu^{ei} = {4.82}\, n\, \ln \Lambda^{ie}_c\, m_e / m_i$. 
The Coulomb logarithms for ion-ion and ion-electron collisions ($\ln \Lambda_c^{ii}$ and $\ln \Lambda_c^{ie}$) are defined as in the NRL Formulary \citep{NRL_Formulary} and discussed by \cite{Fitzpatrick:2015}.}

Specific values for the physical constants are listed in Table \ref{phy-vals.tab}.
These are broadly in line with accepted solar values, with the exception of the the coefficient of dynamic viscosity, which is further modified by a multiplicative factor of $1/(1 + {\alpha}^2)$, where $\alpha = 10^2 \times \lambda^{\rm mfp}_i\, \partial_s \ln A$.
This multiplicative factor guarantees that the dynamic viscosity becomes small as the ion mean-free-path ($\lambda^{\rm mfp}_i = 8.4\times10^3\, T_i^2/n$) approaches or exceeds the geometric length scale, emulating the transition to a collisionless regime as discussed by \cite{Endeve:2001m} \citep[see also the discussion in ][]{Longcope:2010a}.

\begin{table}[ht]
    \centering
    \begin{tabular}{r|r}
            $\gamma$ \hfill \,
            & $5/3$\\
        $m_{i,e} \hfill [g]$ 
            & $2.171\times10^{-24},\, 9.11\times10^{-28}$\\
        $k_B \hfill [\rm erg\, K^{-1}]$ & $1.38\times10^{-16}$\\
        $\mu_i^{(0)} \hfill \left[\rm g\, cm^{-1}\, s^{-1}\right]$ 
            & $2.522\times10^{-15}$\\
        $\kappa_{i,e} \hfill \left[{\rm erg}\, {\rm cm}^{-1}\, \rm s^{-1}\, {\rm K}^{-7/2}\right]$ 
            & $3.2\times10^{-8},\,7.8\times10^{-7}$ \\
    \end{tabular}
    \caption{Values of all physical constants used in fluid model, \rev{expressed in cgs units.}}
    \label{phy-vals.tab}
\end{table}

In addition to the above fluid equations we also calculate the time-dependent ionization of oxygen as a trace element subject to the evolving fluid.
This evolution is decoupled from the energy balance, and does not affect the radiative loss function, which would require tracking additional elements such as He, Fe, S, C, etc., in order to be calculated self-consistently.
The ion charge-state distribution of oxygen, with atomic number $Z=8$, evolves according to 
\begin{equation} \label{ionization.eq}
    D_t Y_j = n_e \left[ {I_{j-1} Y_{j-1} - (I_j + R_j) Y_j + R_{j+1} Y_{j+1}} \right],
\end{equation}
where $D_t \equiv (\partial_t + u\, \partial_s)$ is the material derivative along the flow and $Y_j$ are the individual ionization fractions (i.e., the fraction of a given species \rev{with atomic number} $Z$ having a specific ionization level $j$). $I_j$ and $R_j$ are the temperature-dependent ionization and recombination rates from state $j$, which are calculated as in \cite{Bradshaw:2003a} with updated values from version 8 of the CHIANTI atomic database \citep{Dere:1997, DelZanna:2015}.
Prohibiting ionization and recombination beyond the range of allowed states, the sum of Eq. \eqref{ionization.eq} over all ionization levels vanishes by construction and the system is both normalized and regularized so that
\begin{equation}
    0 \le Y_j \le 1 \qquad \text{and} \qquad \sum_{j=0}^{Z} Y_j = 1.
\end{equation}

\subsection{Steady-state Conditions}

We use the HYDRAD code \citep{Bradshaw:2013} to solve equations \eqref{continuity.eq}--\eqref{ionization.eq} on a numerical grid of 512 cells that are exponentially spaced in heliocentric radius, spanning a range of $0 \le s \le 30 R_\odot$ above the solar surface, where $R_\odot = 700\, \rm Mm$ is the radius of the sun. 
Beyond this base grid we allow up to 16 levels of adaptive refinement through pairwise splitting and merging of cells.
The grid spacing $\Delta s$ is checked against the length scale of the primitive variables ($\lambda_{HD}$) after every tenth timestep in order to maintain the resolution requirement $0.05 \leq \Delta s / \lambda_{HD} \leq 0.1$.
During adaptive refinement the mass, momentum, and energy densities are explicitly conserved to numerical accuracy.
At the limit of refinement the minimum allowable grid spacing is $\Delta s_{\rm min} \approx 10^4{\rm cm}$ at the base of the numerical domain;
however, this extreme resolution is never required in practice, indicating that the transition region is fully resolved at all times.

In order to determine steady-state conditions for the open- and closed-field regions the calculation is first initialized with temperature and density profiles that are generated from a separate routine assuming zero velocity and uniform heating.
The magnetic geometry is spherically symmetric with 
\begin{equation}
    A(s) = (1+s/R_\odot)^2
\end{equation}
and
\begin{equation}
    g(s) = g_\odot \times (1+s/R_\odot)^{-2},
\end{equation}
where $g_\odot = -2.74\times10^4{\rm \,cm}{\,\rm s}^{-2}$ is the gravitational acceleration at the solar surface.
The cross-sectional area is dimensionless and normalized to unity at the solar surface.
The density and temperature are set to $T_0 = 2\times10^4{\rm \,K}$ and $n_0=10^{10}{\,\rm cm}^{-3}$ at a height of $10^{9}{\rm cm}$ above the solar surface, which sets the location of the base of the transition region within the domain. 
Above this height the temperatures of the ions and electrons rise abruptly through the transition region to coronal values, while below this level the temperature is effectively uniform and the density increases exponentially toward the interior of the solar surface.

At the beginning of the simulation the initially-uniform heating is replaced with a superposition of three exponentials of the form $h_i(s) = h_0 \exp(-s/s_l)$, after which the system is advanced in time and allowed to relax toward a new steady-state solution.
\rev{Exponential heating functions have been used extensively in solar wind models \citep[see, e.g., ][]{Pinto:2009a} and are employed here for simplicity; although in the future we intend to incorporate more realistic mechanisms \cite[such as those discussed by][]{Cranmer:2007}.}
Specific values for the three ion heating rates and scale heights are given in Table \ref{heating.tab}.
The electrons are not heated directly ($h_e = 0$), but instead rely on collisional coupling in the lower corona and thermal conduction in the extended corona to maintain their temperature.
This particular heating model serves to give realistic values for the electron temperature in the lower-mid corona and for the mass flux (per unit solid angle) into the heliosphere.

\begin{table}[ht]
    \centering
    \begin{tabular}{l c c}
    Deposition Region      & $h_0 [{\rm erg}{\,\rm cm}^{-3}{\,\rm s}^{-1}]$  & $s_l [{\rm cm}]$ \\ \hline
    Transition Region      & $2.9\times10^{-5}$ & $7.0\times10^9$ \\
    Lower Corona           & $8.7\times10^{-7}$ & $2.1\times10^{10}$ \\
    Extended Corona        & $5.0\times10^{-9}$ & $7.0\times10^{10}$
    \end{tabular}
    \caption{Heating parameters for ion energy deposition. These are used for both the hydrostatic and transonic steady-state conditions and the dynamic reconnection runs. Electron heating is set to $h_e = 0$ in all cases.}
    \label{heating.tab}
\end{table}

\begin{figure}[ht]
    \centering
    \includegraphics[width=\linewidth]{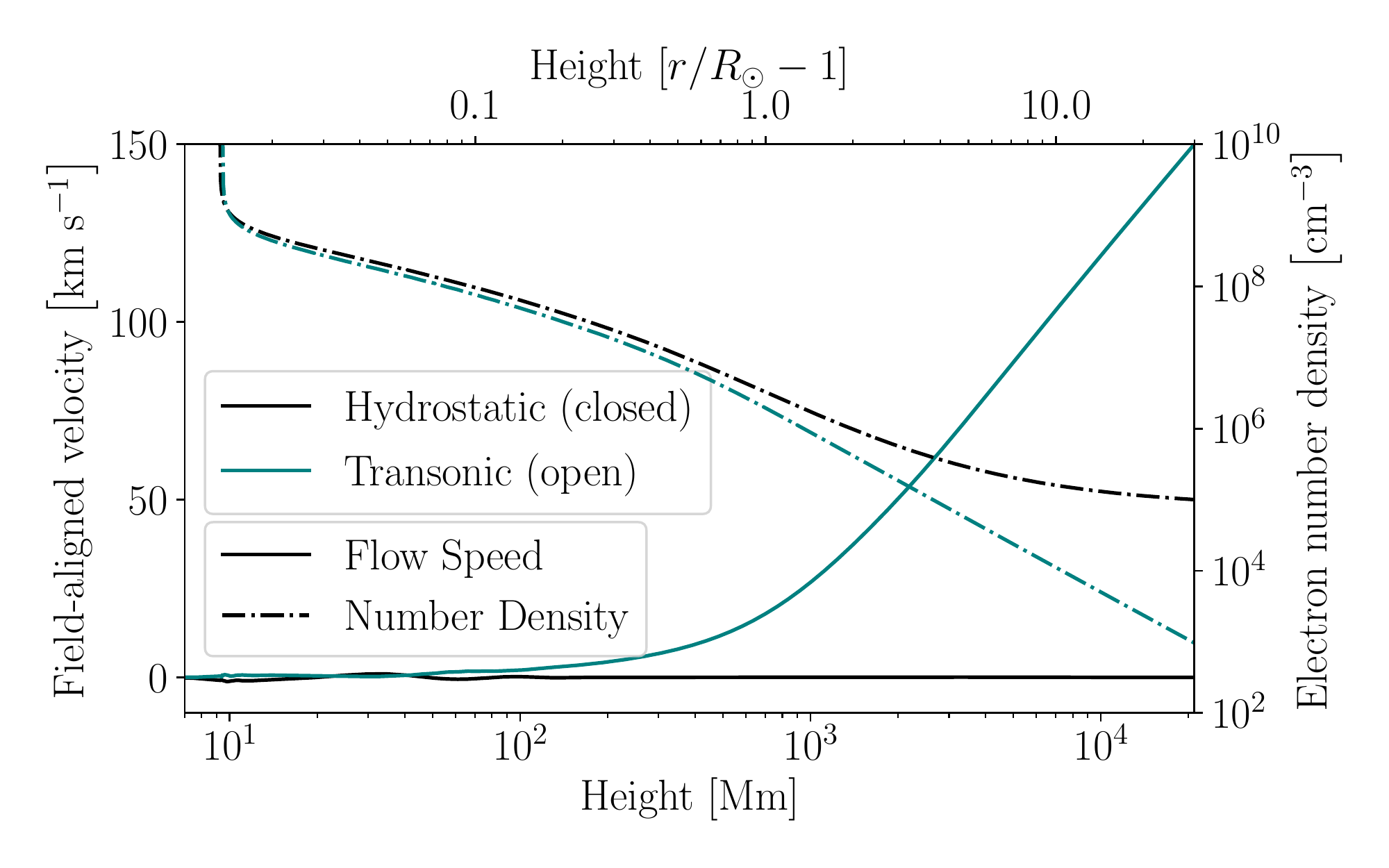}\\
    \includegraphics[width=\linewidth]{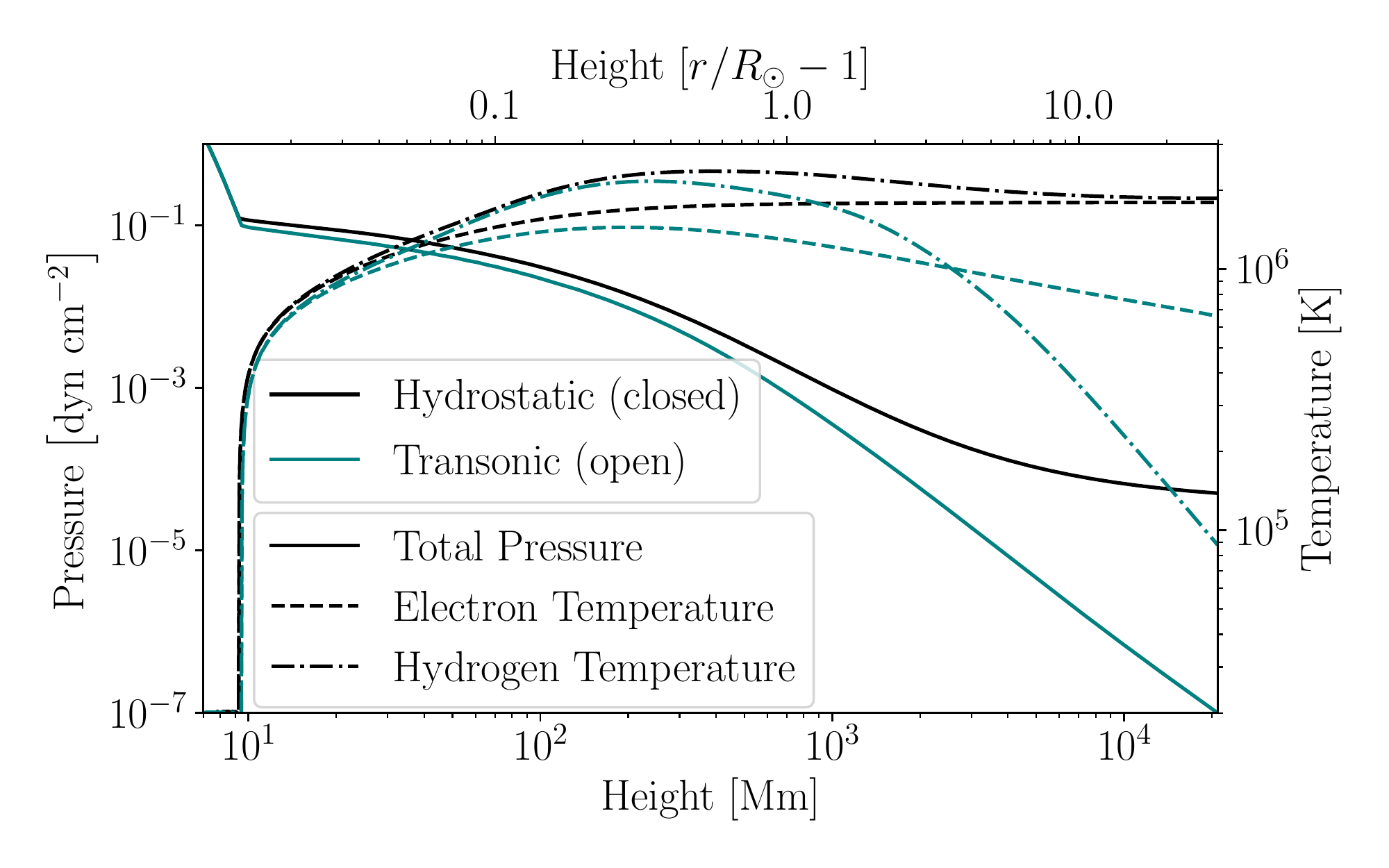}
    \caption{Transonic (teal) and hydrostatic (black) relaxed states calculated using HYDRAD with open and closed boundary conditions at $s=30R_\odot$. In the upper panel the flow speed is indicated by the solid curves while the particle number density is shown in the dash-dotted curves. In the lower panel the total pressure (sum of ion and electron pressures) is shown by the solid curves while the individual ion and electron temperatures are shown by the dashed and dash-dotted curves. The base of the transition region is clearly visible at a height of $\lesssim 10 \rm \, Mm$, above which the temperatures rise from $T_0 = 2 \times 10^4 K$ to coronal values on the order of $10^6 \rm \, K$.}
    \label{fig.initial_hyd}
\end{figure}

The initial relaxation is performed subject to two sets of boundary conditions at the radial outer limit of $s_{\rm max} = 30 R_\odot$.
In the case of set 1: the mass, momentum, and energy densities are linearly extrapolated across the boundary, and the pressure is systematically reduced in the ghost cells outside of the domain in order to encourage the development of a supersonic outflow.
Once this outflow has been achieved the pressure reduction is removed and the temperature gradient is adjusted to prevent the formation of inward heat fluxes across the boundary.
Following the transition of the outflow to a supersonic condition, a sonic point forms within the domain and migrates progressively inward until it stabilizes in the lower-middle corona \rev{at a height of roughly $7\, R_\odot$.
The ultimate location of the sonic point is somewhat higher than might be expected owing to the fact that in our model the wind is accelerated entirely by the thermal pressure gradient, which results in somewhat slower wind speeds than if momentum is injected directly \citep{Holzer:1982}.}
Because the typical flow speed is of the same order as the sound speed in this configuration, the system achieves a quasi-steady transonic wind solution on the timescale of the acoustic travel time across the domain, which is of order $10^4{\rm s}$.
This quasi-steady transonic solution is depicted by the teal curves in Figure \ref{fig.initial_hyd}.

In the case of set 2: the mass, momentum, energy, and heat fluxes are all fixed to zero through the outer boundary, forcing the system to be closed.
This causes the solution to settle toward a hydrostatic state in asymptotic time; because the slowest mode of the system is governed by the transport of mass, the relaxation time scales as the size of the domain divided by the fluid speed, which diverges as the fluid velocity approaches zero.
The quasi-steady hydrostatic solution is depicted by the black curves in Figure \ref{fig.initial_hyd}.
While the system never fully achieves a hydrostatic state, we consider it to be sufficiently relaxed when the maximum value of the coronal mass flux has fallen below $1\%$ of the value of the transonic wind solution at the same location.
This requirement is significantly more strict than the usual hydrostatic scaling of $\rho u^2 \ll P$ (i.e., small Mach number), and is enforced specifically to ensure that the velocities of the hydrostatic and transonic solutions are well separated in the lower corona where even the transonic solution is significantly subsonic.

The equilibration time of the hydrostatic solution is extremely long (approximately $10^8{\,\rm s}$) owing to our strict relaxation criteria and the length of the numerical domain, which is larger than any closed-field/hydrostatic configuration that might be found in the solar environment.
The reason for constructing such a solution is to ensure compatibility between the numerical domains of the open- and closed-field configurations, and since it is only the lower portion of the plasma column that is needed from the closed-field initial condition, the unphysical radial extent of this solution is not a significant concern.
Were the hydrostatic initial condition to be extracted from a curved field line with its apex in the lower or middle corona, the average volumetric heating rate along the entire field line would be increased, with the energy being distributed within a smaller volume, and the solution would likely be hotter and more dense, thereby exaggerating the differences already present in these two calculations.

For both of the solution profiles depicted in Figure \ref{fig.initial_hyd} the location of the base of the transition region (where $T_e$ and $T_i$ begin to rise abruptly from their minimum values of $2\times10^4 \,\rm K$) has fallen by a few hundred $\rm km$ from its initial location of $10\rm Mm$, with the final location being slightly lower for the hydrostatic solution than for the transonic solution.
This settling is expected as the structure and location of the transition region are determined by the evolving mass and energy fluxes through the model chromosphere and lower corona, which change in time as the system relaxes toward an equilibrium state.
Beyond this, the structure of the transition region and lower corona is very similar in both cases, with temperatures and densities that closely agree up to a height of $\sim 20 \,\rm Mm$ above the transition region.
Both solutions also exhibit oscillatory behavior in the velocity profiles up to a height of $\sim 100 \, \rm Mm$, above which height the transonic solution then begins to accelerate quickly into the middle-corona.

\subsection{Dynamic Equilibria of Oxygen Ionization Levels}

During the initial relaxation the ionization levels are not computed as these add substantial overhead to the calculation.
Following the equilibration of the fluid profiles, we then initialize the ionization levels from a set of previously calculated equilibrium ionization solutions, which are parameterized in electron temperature and interpolated onto the numerical domain using the local plasma temperature.
The two simulations are then advanced in time until the time-dependent ionization again converges to a steady state in each case.

\begin{figure}[ht]
    \centering
    \includegraphics[width=\linewidth]{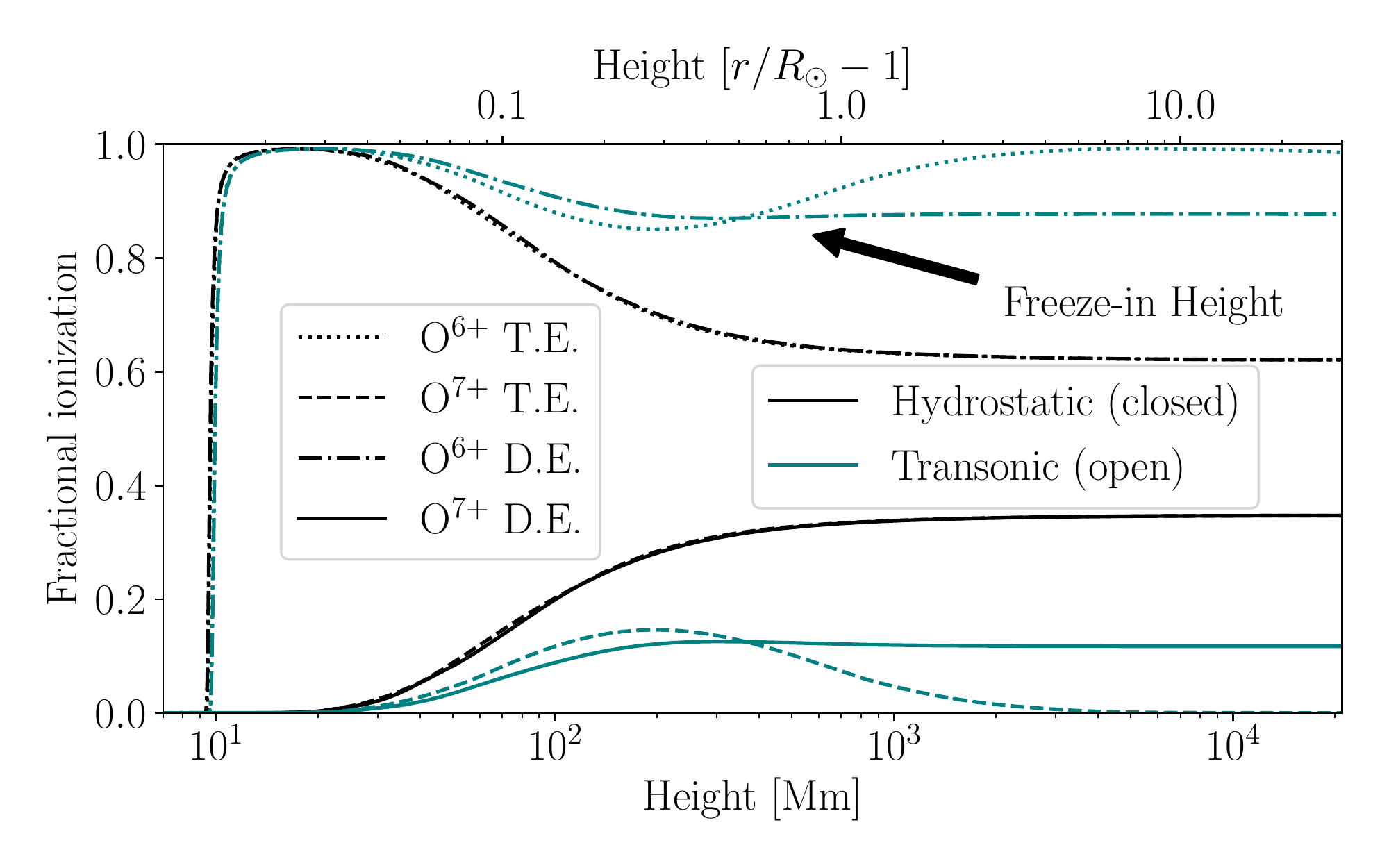}
    \caption{Ionization fractions $\rm O^{6+}$ and $\rm O^{7+}$ for hydrostatic and transonic solutions assuming either a thermal equilibrium (T.E.) or dynamic equilibrium (D.E.). For the hydrostatic case the two solutions are nearly identical, while the dynamic equilibrium for the transonic case is a clear departure from the thermal equilibrium solution, especially near and above the freeze-in height ($s \geq h_f \sim 0.5 R_\odot$).}
    \label{fig.initial_nei}
\end{figure}

We refer to the relaxed ionization profiles as ``dynamic equilibria'', to distinguish them from the assumed ``thermal equilibrium'' solutions that explicitly ignore advective transport.
For the hydrostatic case the two solutions are nearly identical since there is no significant flow so the solution is characterized by an instantaneous balance of all ionization and recombination rates with 
\begin{equation}
    \lim_{u \to 0} \left[ {I_{j-1} Y_{j-1} - (I_j + R_j) Y_j + R_{j+1} Y_{j+1}} \right] \to 0.
\end{equation}
For the transonic solution, however, the dynamic equilibrium is set by a competition between the instantaneous temperature of the fluid and the advective transport, which informs the temperature history of the fluid and, hence, the rate equations.
The dynamic equilibrium ionization is characterized by 
\begin{equation} 
\partial_s Y_j = \left[ {I_{j-1} Y_{j-1} - (I_j + R_j) Y_j + R_{j+1} Y_{j+1}} \right] / u,
\end{equation}
which exhibits spatially uniform ionization levels above the so-called ``freeze-in'' height ($h_f$), where the ionization and recombination timescale ($\tau_j$) becomes long compared to the travel time across the temperature scale height so that 
\begin{equation} 
\partial_s Y_j \sim Y_j / (\tau_j u) \rightarrow 0.
\end{equation}
In principle this freeze-in effect occurs separately for each ionization level; however, in practice the zero-sum nature of the rate equations and the limited number of ionization states with non-zero populations in the corona causes the effect to occur coherently across these states.
In the following sections we will refer to the ionization fractions $Y_j$ by their respective ionization states $\rm O^{j+}$, with the understanding that these correspond to the underlying spatial-temporal populations.
The final, quasi-steady states of both the open- and closed-field solutions are shown in Figure \ref{fig.initial_nei}, with both the initial (assumed)  thermal equilibrium and relaxed dynamic equilibrium ionization levels shown for comparison.

\subsection{Reconnection Model}

\begin{figure}[ht]
    \centering
    \includegraphics[width=\linewidth]{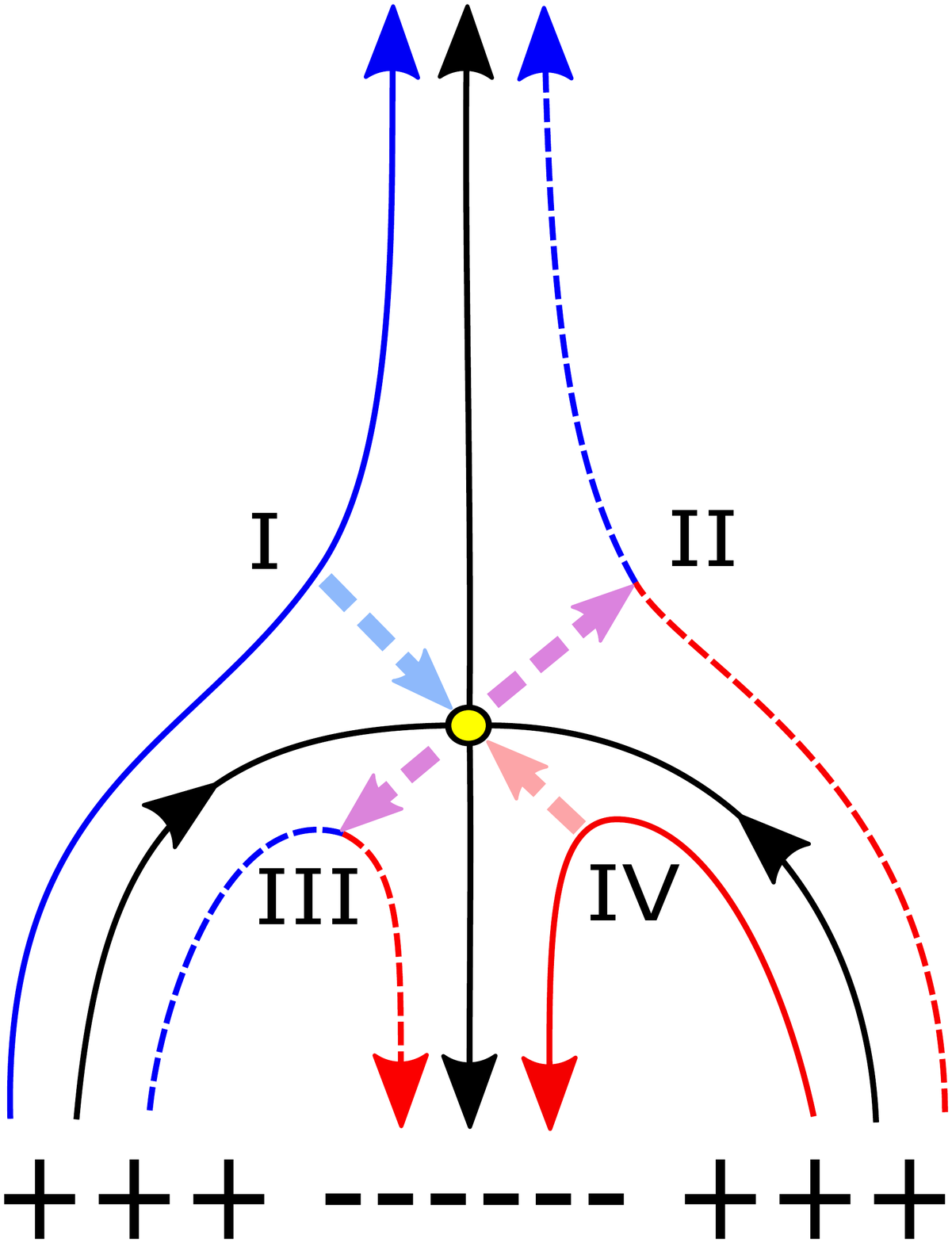}
    \caption{Schematic of interchange reconnection through an isolated coronal null point (yellow) above a region of parasitic polarity on the solar surface. Separator lines (black) emanate from the null point and partition the surrounding volume into four magnetic flux domains, two that are open and two that are closed. Field lines enter the reconnection site from the inflow domains (I and IV) and recede into the outflow domains (II and III). Newly reconnected field lines (dashed) are comprised of segments of pre-reconnection field lines (solid) that originated from within adjacent domains, as indicated by their coloring (red/blue).}
    \label{fig.rxn}
\end{figure}

While magnetic reconnection in three dimensions (3D) can occur throughout a distributed non-ideal region \citep{Priest:2003ja}, the behavior that we have in mind for this study is the more discrete null-point reconnection \citep{Pontin:2013}, in which the magnetic diffusivity is assumed to be negligible everywhere except within a very small volume surrounding an isolated coronal null point.
A schematic of this process is shown in Figure \ref{fig.rxn}, which depicts a 2D flux system with four magnetic domains that overlie a region of negative polarity flux within a larger region of positive polarity flux.
This geometry is representative of a slice through a coronal pseusostreamer, as might be found above an isolated magnetic bipole within a larger unipolar region on the solar surface.

Two of the magnetic domains (I and II) are open to the heliosphere while the other two (III and IV) are closed, mapping from positive to negative polarity regions on the solar surface.
The black field lines that partition these domains represent separatrices (or separatrix surfaces in 3D), which emanate from the magnetic null point (yellow dot).
During interchange reconnection field lines from domains I and IV are swept into the null point where they reconnect and subsequently retract into domains II and III.
As magnetic flux is processed through the reconnection site the plasma is similarly swept across the domain boundaries so that immediately after reconnection the plasma properties along each segment of the reconnected field lines are inherited from their respective source regions.

We construct the post-reconnection initial condition along a representative field line from segments of steady-state conditions for both open- and closed-field configurations, which we concatenate to form a piecewise steady state with a discontinuity at the reconnection site.
This construction assumes that plasma in the inflow domains remains in a steady state, even as field lines are swept toward the reconnection site, so the tacit assumption is that the reconnection rate is sufficiently small for motion of the field lines themselves to have no effect on the field-aligned dynamics.
Clearly this assumption ignores certain important aspects of null-point reconnection, including the expansion of the flux-tube cross section in the vicinity of the null-point;
however, by suppressing these effects we are able to study the fluid-driven evolution more closely and with fewer confounding influences.

From the post-reconnection initial condition we again use the HYDRAD code to evolve the system in time.
To ensure a monotonic transition between the two fluid states on either side of the reconnection site, the constructed post-reconnection state is initially refined using linear interpolation and suppressing explicit conservation of the mass, momentum, and energy densities.
Following this initial refinement, the subsequent integration is performed exactly as described in Section \ref{sec.fluid}.
We perform six calculations in total, using the same heating profiles as in the steady state calculations, with the reconnection site placed at $H_r \in \left \{R_\odot/8, R_\odot/4, R_\odot/2, R_\odot, 2 R_\odot, 4 R_\odot \right \}$.
Time-dependent ionization of oxygen is tracked in addition to the hydrodynamic evolution and state-files are output every $100 \, \rm s$ of model time.
The total simulation model time is $4 \times 10^5 \, \rm s$ (about 100 hours) from the instant of reconnection in each case, sufficient for the fluid to settle to a quasi-steady state that is indistinguishable from the transonic initial condition.

\section{Results}\label{results.sec}

\subsection{Initial Riemann Decomposition}

Because the numerical domains and magnetic geometries ($s$, $g$, and $A$) are identical between both the open- and closed-field steady-states, the post-reconnection initial condition begins in equilibrium everywhere except at the reconnection site.
There, the initial discontinuity subsequently evolves as a Riemann problem, decomposing into a shock and a rarefaction wave (as well as ion and electron thermal fronts), as needed for the jump conditions across each feature to collectively describe the total change in each of the fluid variables across the discontinuity.
In particular, since the wind solution above the reconnection site exhibits a flow that is directed away from the hydrostatic condition below the reconnection site, the collective response is that of a rarefaction; however, because the wind is of lower density than the hydrostatic solution, any outward propagating feature must be compressive.
The Riemann solution is, therefore, composed of a leading shock, which propagates outward at supersonic speed in the rest frame of the expanding wind, combined with a rarefaction wave whose leading edge propagates inward at the sound speed in the rest frame of the hydrostatic column below the reconnection site.

\begin{figure}[ht]
    \centering
    \includegraphics[width=\linewidth]{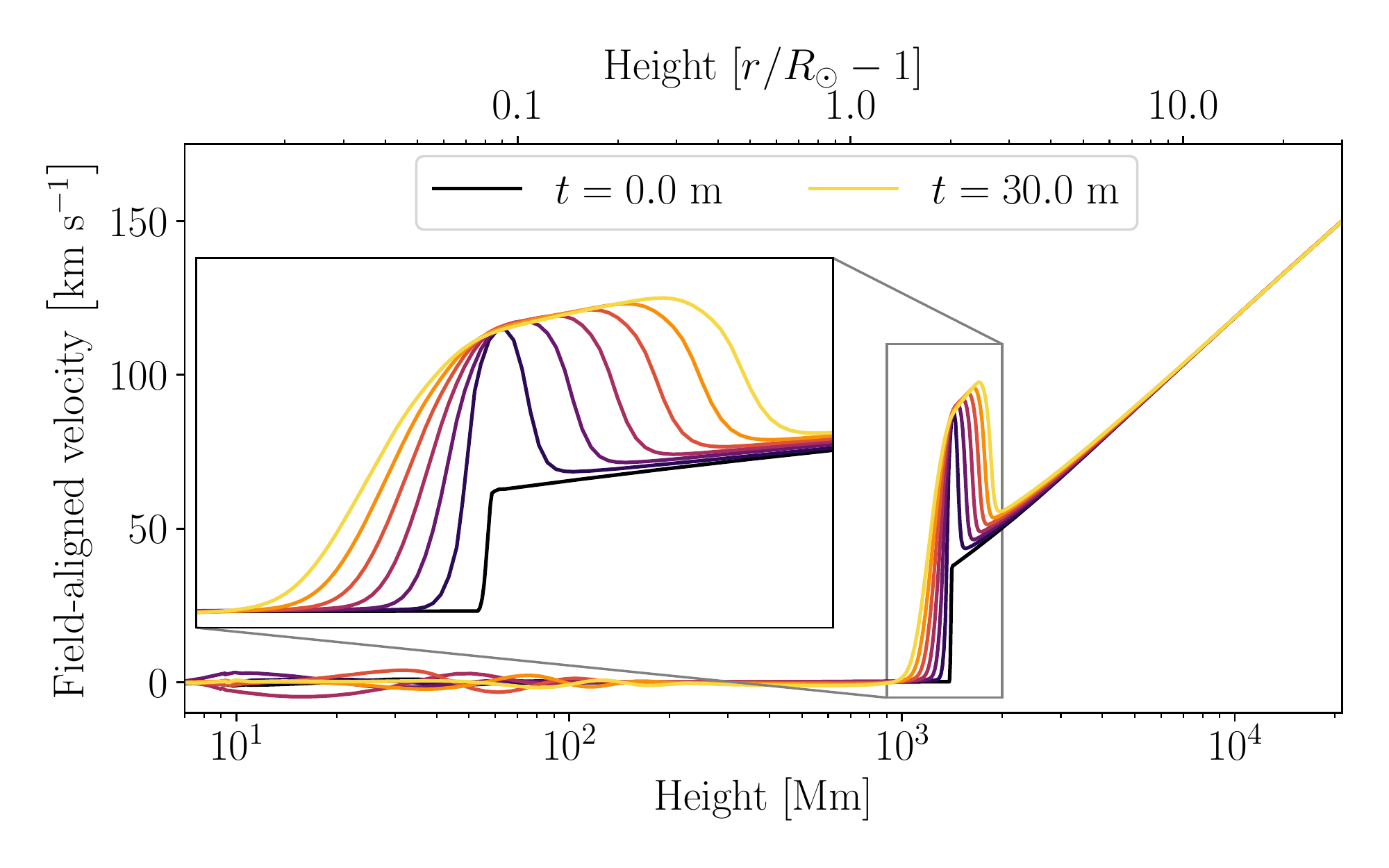}\\
    \includegraphics[width=\linewidth]{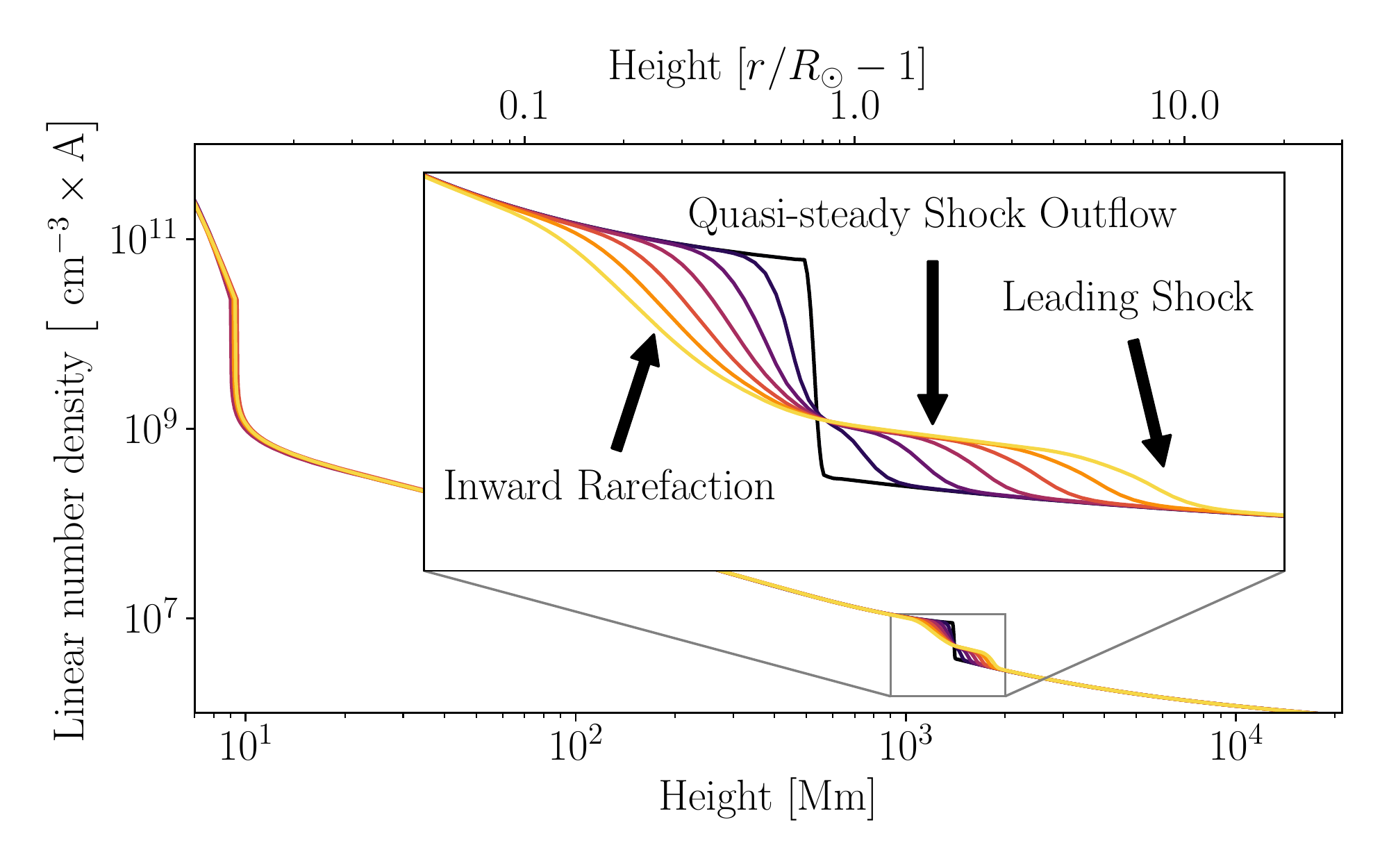}\\
    \includegraphics[width=\linewidth]{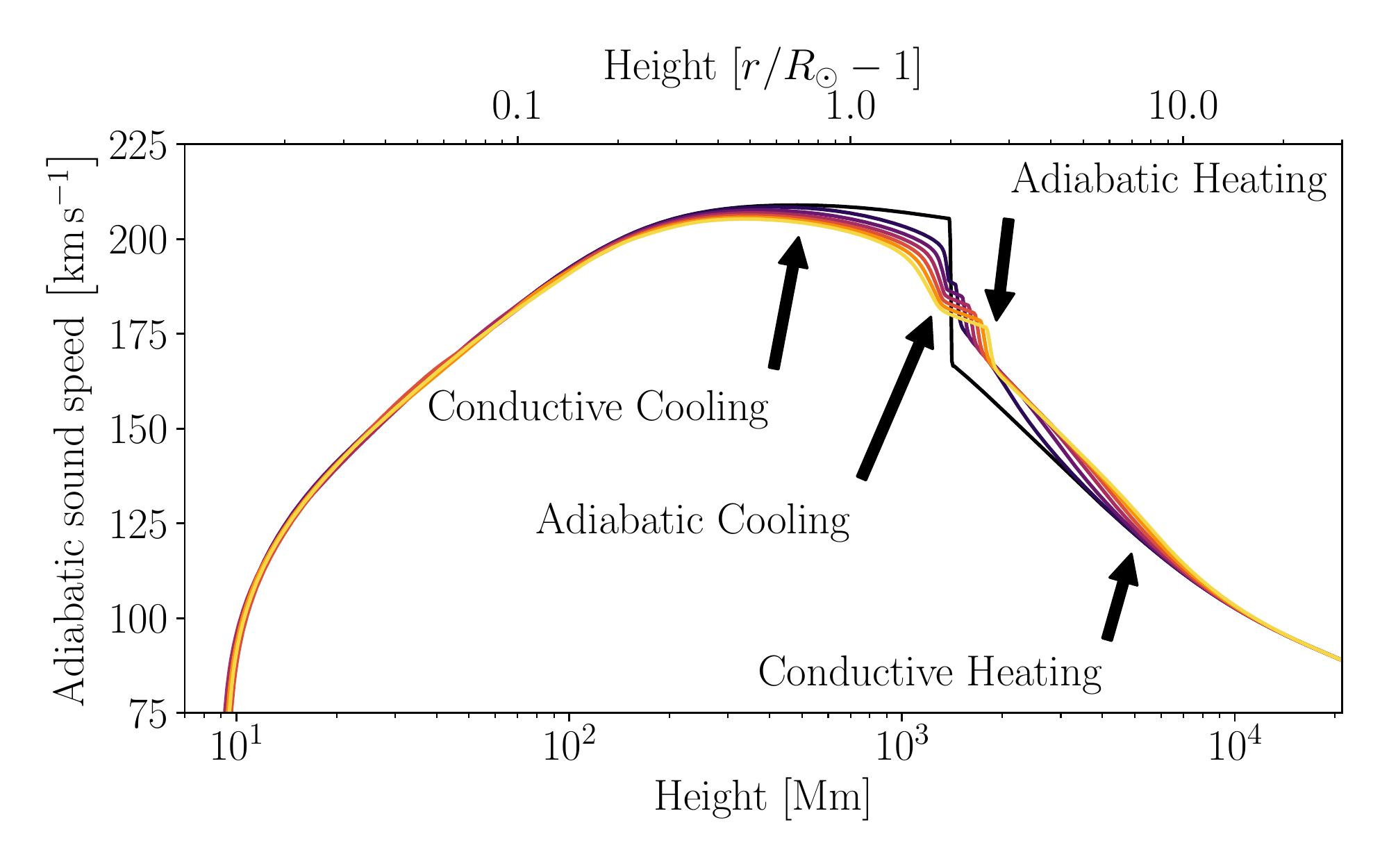}      
    \caption{Velocity, linear number density, and sound speed of the evolving Riemann solution for a reconnection event at $H_r = 2 R_\odot$. The shock and rarefaction propagate away from the initial discontinuity, and are visible in the velocity and density profiles. The sound speed reflects the mean temperature of the two species, and exhibits dynamics resulting from both the hydrodynamic evolution as well as electron thermal conduction.}
    \label{fig.riemann}
\end{figure}

This initial evolution is depicted in Figure \ref{fig.riemann}, for a reconnection event located at $H_r = 2 R_\odot$.
The outward propagating shock and inward rarefaction are clearly visible in the velocity, which increases with height across the rarefaction and then decreases again across the shock, while the linear number density ($n A$) decreases with height across the rarefaction and then decreases again across the shock. 
Note that in the rest frame of the shock the fluid appears to be moving inward, so the low-speed/high-density post-shock region lies between the shock and the rarefaction, and this region grows in time as the two features propagate away from each other.
As the rarefaction propagates inward it grows in time but continues to connect the quasi-static region below to the post-shock region above, so that while the leading edge propagates downward at the acoustic speed, the trailing edge remains nearly fixed at approximately the location of the initial discontinuity, which behaves as a stationary point for both the velocity and number density during this initial evolution.

The temperatures of the two species behave similarly to the number density during the initial Riemann evolution, but with additional dynamics resulting from viscous heating,  thermal conduction, and compressive (adiabatic) heating and cooling.
Because the thermal conductivity of the ions is relatively weak, their temperature evolves primarily in response to compressive and advective transport, so it tracks closely with the number density profile over time.
However, the highly efficient electron thermal conductivity quickly dissipates the initial structure within the electron temperature profile creating broad heating and cooling fronts that extend far ahead of the shock and rarefaction, cooling the hydrostatic column below the reconnection site and heating the expanding wind.
These temperature variations are visible in the adiabatic sound speed $c_s^2 = \gamma (p_e + p_i) / \rho $ (which reflects the average temperature of the two species), as depicted in the lower panel of Figure \ref{fig.riemann}.

\subsection{Rarefaction Reversal}

\begin{figure}[ht]
    \centering
    \includegraphics[width=\linewidth]{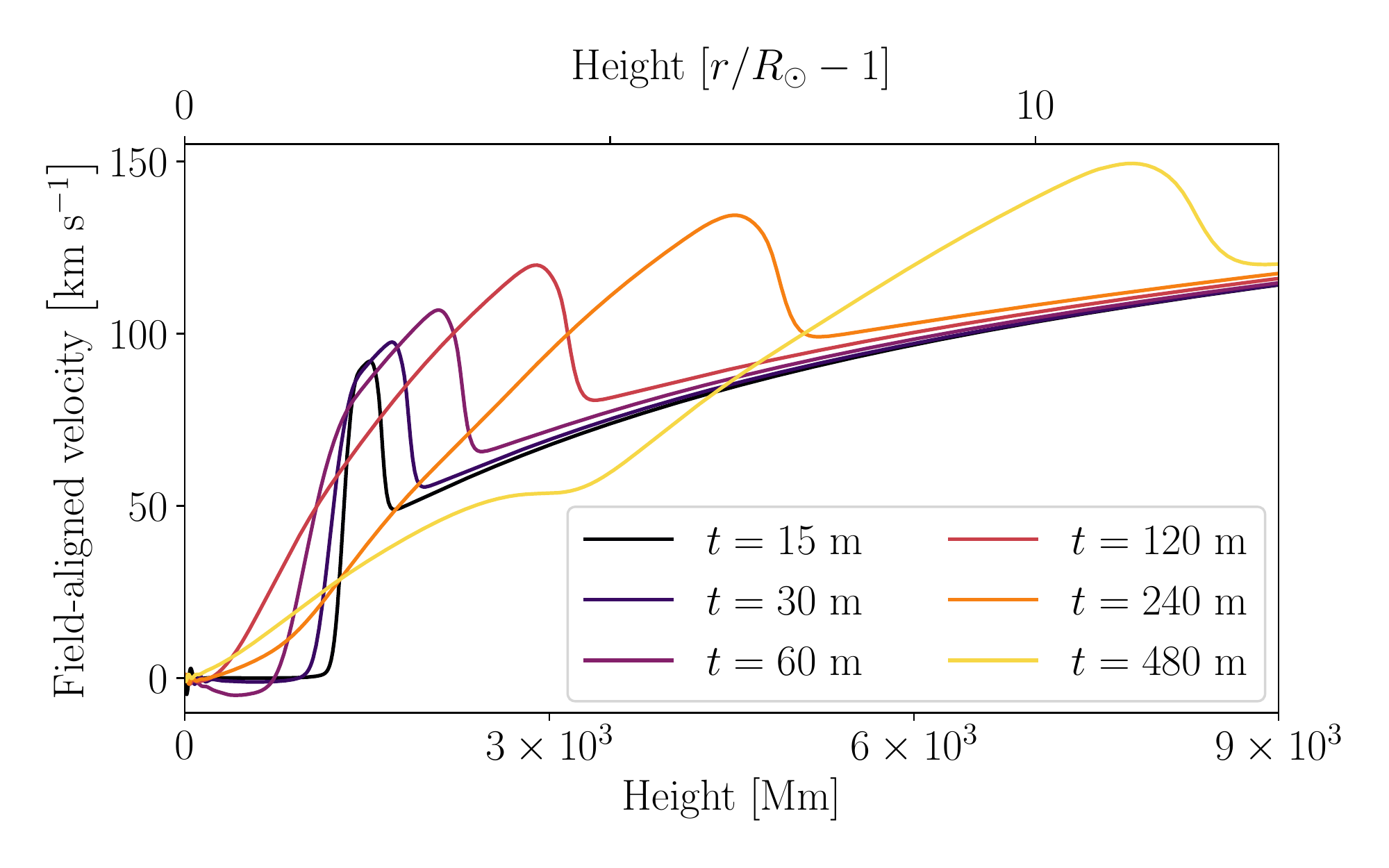}\\
    \includegraphics[width=\linewidth]{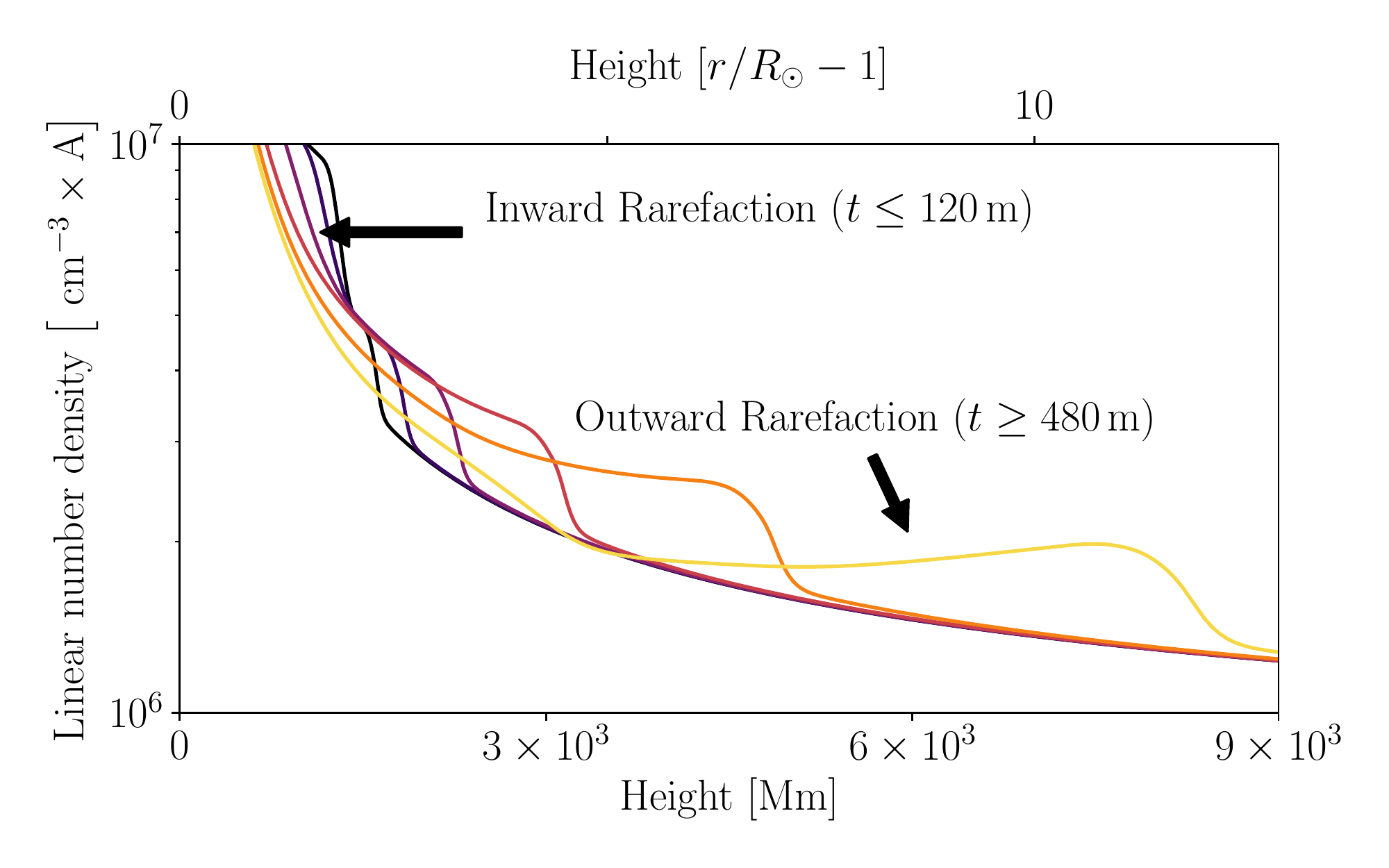}
    \caption{Velocity and linear number density during reversal of the rarefaction. Initially, the linear number density decreases with height through the rarefaction, consistent with inward-propagation; however, after reversal of the rarefaction the leading edge propagates outward and overtakes the leading shock, and the linear number density within the rarefaction increases with height.}
    \label{fig.reversal}
\end{figure}

As the rarefaction propagates inward, the temperature below it is preconditioned by the reduction of the downward heat flux in the hydrostatic region below the reconnection site. 
This causes a weak downflow to develop in the region between the transition region and the rarefaction, which is quickly subsumed by the accelerating outflow across the leading edge of the rarefaction.
This is visible in Figure \ref{fig.reversal} between $t=30\,\rm m$ and $t=60\,\rm m$.
Meanwhile, an outward mass-flux begins to develop through the transition region as the reduced temperature leads to a reduction in radiative cooling, which therefore requires an outward enthalpy flux $(E + P) u$ in order to balance the local heating rate.
Eventually the leading edge of the rarefaction stalls against the up-welling of mass through the transition region as it seeks a new equilibrium that is consistent with the electron temperature profile, which settles rapidly toward a wind-like solution.
As the leading edge of the rarefaction stalls, the interior continues to expand over time, and the upper extent of the rarefaction accelerates outward, eventually becoming identifiable as the leading edge of a newly formed (reflected) outward-propagating rarefaction.

The reversal of the rarefaction coincides with a reversal in the linear number-density profile, which subsequently increases with height (see discussion in the Appendix), and cannot (in the absence of a second shock) match the density jump required to connect the post-shock outflow to the steadily decreasing density at the top of the transition region, which continues to fall as it seeks a wind-like solution.
This effect, combined with the steadily increasing outward mass flux through the transition region, leads to the development of another (weak) trailing shock, which develops behind the now-outward-propagating rarefaction.
Meanwhile, the leading edge of the rarefaction propagates outward at the sound speed in the post-shock medium, which is itself subsonic in the rest frame of the leading shock.
Eventually the outward-propagating rarefaction overtakes the leading shock, and the three features combine to form a single, coherent structure, known as an N-wave\footnote{Note that for leftward propagation the profile is that of an ``N'' while for rightward propagation the structure is inverted.} \citep[this being the ordered combination of two shocks connected by an interior rarefaction, as discussed by][]{Friedrichs:1948}.

The various stages of this reversal are depicted in Figure \ref{fig.reversal}: the downward propagation of the left side of the rarefaction is clearly visible between $t=15\rm m$ and $t=60\, \rm m$, along with the monotonic decrease in the linear number density with height. 
From $t=120\, \rm m$ to $t=240\, \rm m$ the rarefaction accelerates upward, subsuming the post-shock outflow region below the shock, and the linear number density profile begins to flatten.
Eventually, by $t=480\, \rm m$, the upper extent of the rarefaction has overtaken the shock and the linear number density within the rarefaction now decreases inwardly away from the shock.
This reversal in the linear number density is the earliest signature of the newly-formed N-Wave, which is characterized by velocity and density profiles whose gradients are codirectional through the entire structure.

\subsection{N-Wave Structure and Dynamics}

\begin{figure}[ht]
    \centering
    \includegraphics[width=\linewidth]{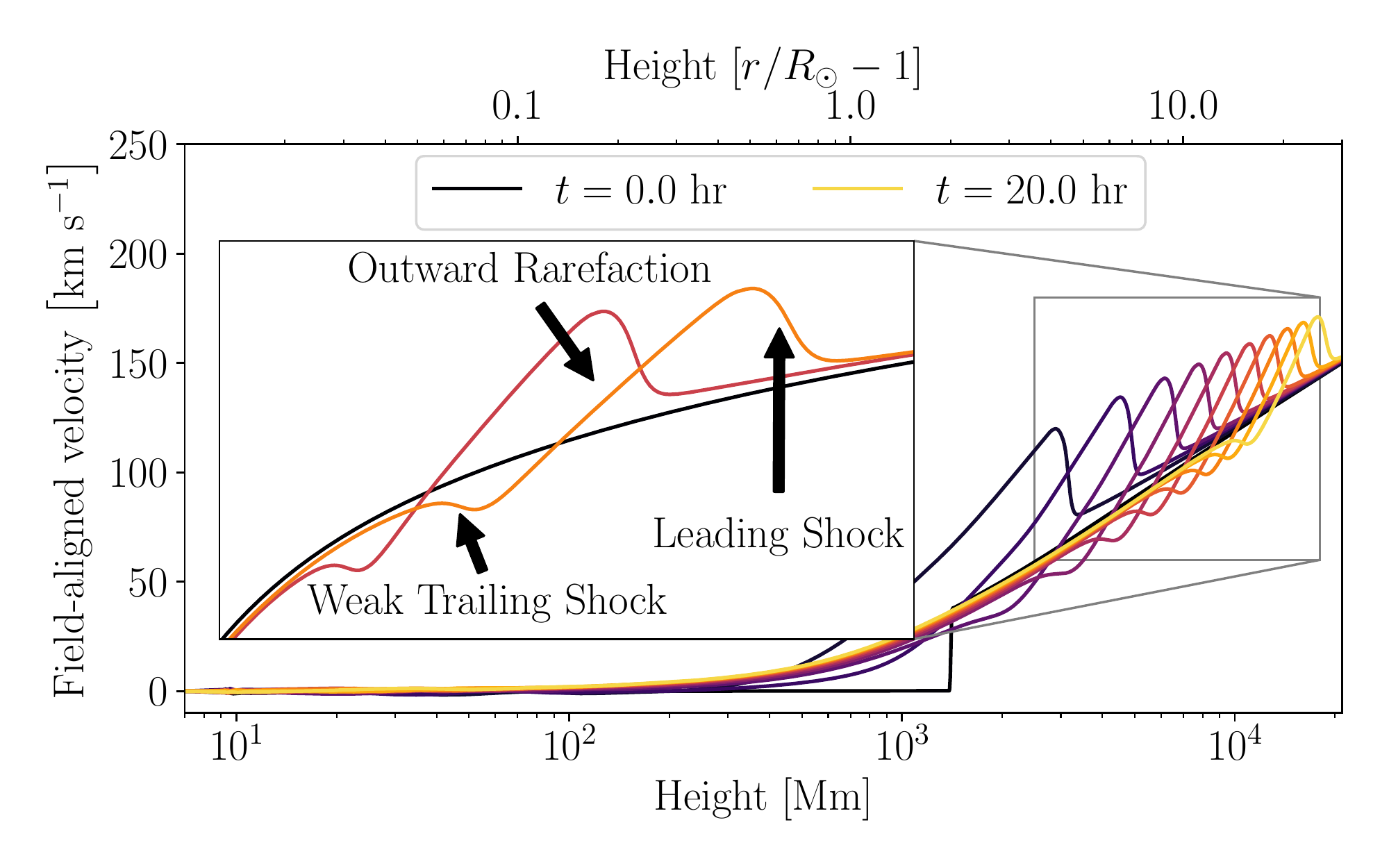}\\
    \includegraphics[width=\linewidth]{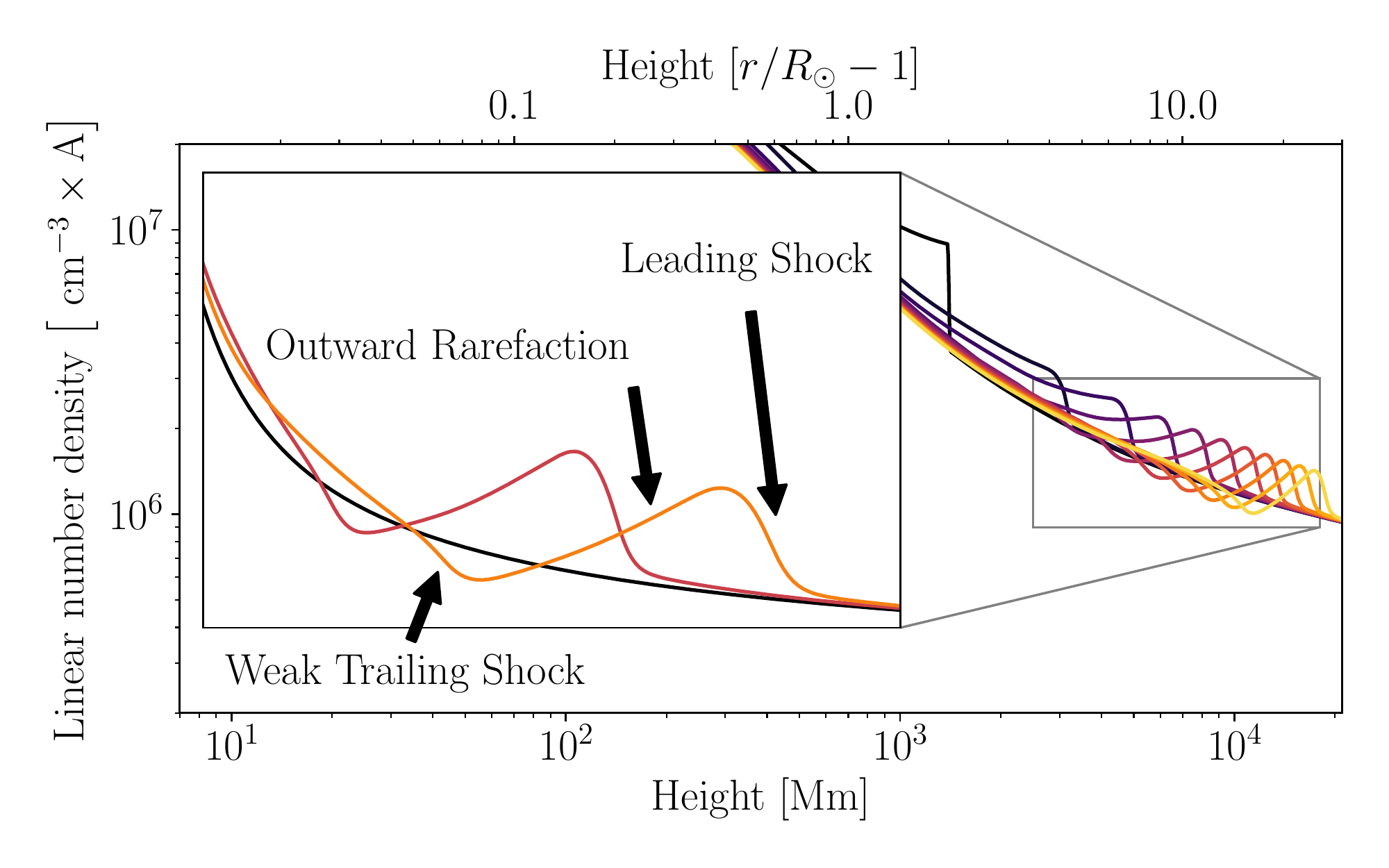}\\
    \includegraphics[width=\linewidth]{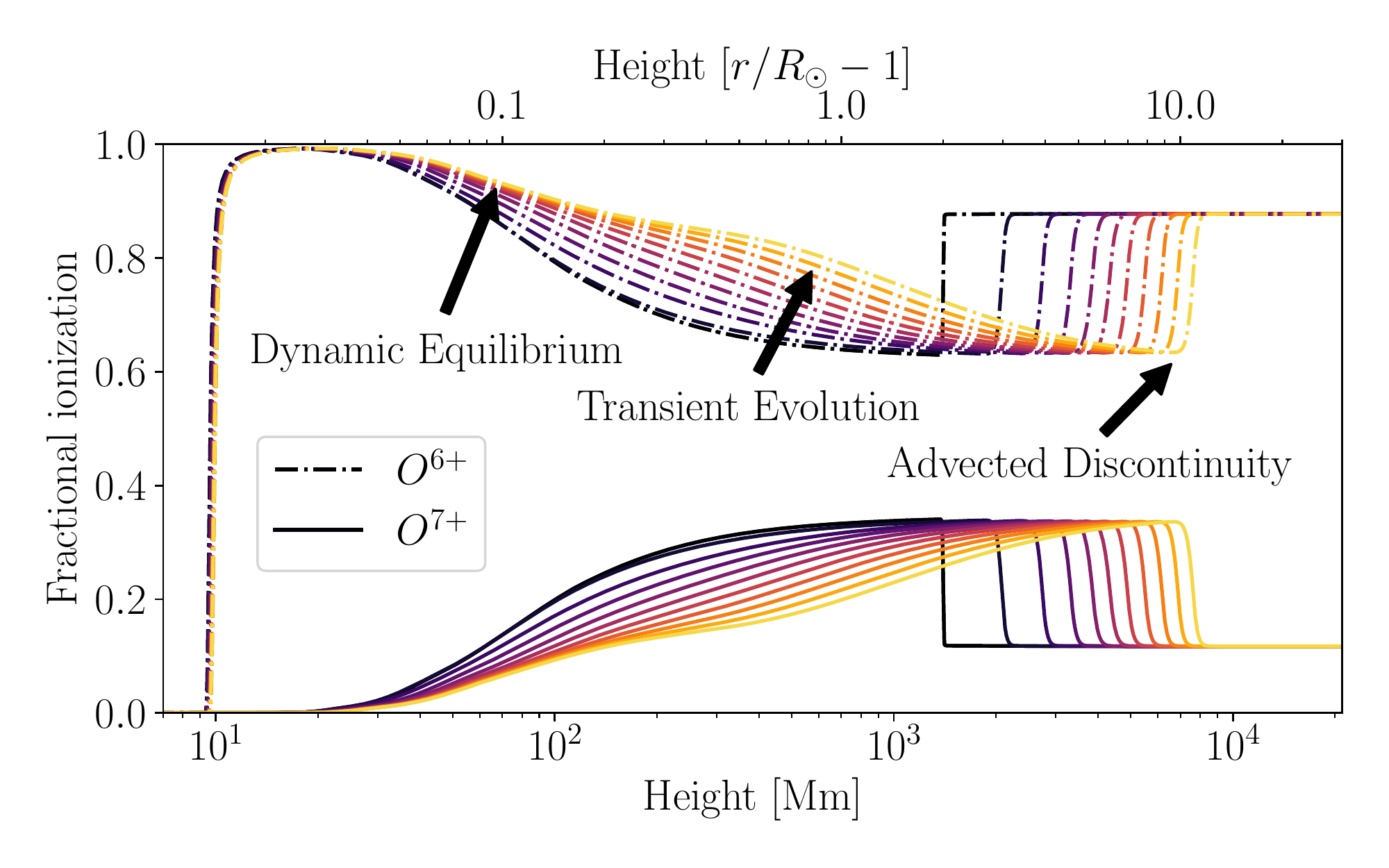}
    \caption{Velocity, linear number density, and selected oxygen ionization fractions $\rm O^{6+}$ and $\rm O^{7+}$ for the propagating N-wave solution. The velocity and density profiles mimic each other within the N-wave, which is composed of a leading shock, an outward-propagating rarefaction, and a trailing shock. The evolution of the ionization profiles reflects a competition between advective transport and spatially-varying ionization and recombination rates, which depend on the evolving temperature profiles.}
    \label{fig.Nwave}
\end{figure}

The structure and evolution of the outward-propagating N-Wave are depicted in Figure \ref{fig.Nwave}, which shows the time-evolution of the fluid velocity, linear number density, and fractional ionization populations $O^{6+}$ and $O^{7+}$.
Note again that while $n$ (not shown) decreases monotonically with height, $n A$ necessarily increases with height in the region between the trailing shock and the leading shock, opposite the behavior of the initially-inward propagating rarefaction shown in Figure \ref{fig.riemann}.
The connection of the leading shock, through the rarefaction, to the trailing shock is also visible in the velocity profiles, which exhibits the eponymous ``N'' shape, decreasing with height across the trailing shock, then increasing quasi-linearly through the rarefaction to a maximum at the leading shock, before then decreasing abruptly across the leading shock onto the pre-shock wind solution.

The ionization fractions, on the other hand, do not participate in the compressive and expansive dynamics of the shocks and rarefaction, despite the fact that the initial discontinuity in the values of $Y_j$ is co-spatial with the initial Riemann problem. 
This follows from the form of the advective component of the rate equation (i.e., the material derivative $u\, \partial_s Y_j$) which causes the ionization states to evolve as passive scalars in the absence of thermal effects (when the ionization and recombination rates become small) so that the ionization fractions are simply carried along by the flow but do not increase or decrease as the fluid expands. 
Accordingly, the initial discontinuity (which is above the freeze-in height in the case of the $H_r = 2 R_\odot$ example) is carried along by the high-speed outflow within the rarefaction, but lags behind the leading shock, which propagates faster than the bulk flow. 
Thus, while the shock reaches the outermost boundary of $30R_\odot$ in just a bit less than $24\,\rm hr$ of simulation time, the ionization signature has only just passed $10 R_\odot$ in that time, and does not arrive at the outer boundary until significantly later.

Below the freeze-in height, the fractional ionization populations of the initially hydrostatic fluid evolve in response to the changing temperature that occurs in both time and space as the fluid begins to expand within the rarefaction and eventually settles onto a new wind solution.
For fluid parcels that originate sufficiently low in the corona, the transit time to the freeze-in height is long compared to both the timescales of fluid equilibration and ionization and recombination, so that by the time the fluid reaches the freeze-in height, the ionization ratios are indistinguishable from a steady-state wind solution.
Fluid that originates higher up within the hydrostatic column has a thermal history that is more strongly representative of the conditions in the closed corona, having less time to equilibrate before it reaches the freeze-in height.
The structure of the ionization profiles following reconnection is therefore given by an abrupt transition from the dynamic equilibrium of the initial wind solution just above the reconnection site to something resembling the dynamic equilibrium of the hydrostatic solution at the same height, followed by a smooth transition back to the quasi-steady wind profile, all of which travels outward at the local fluid speed.

\section{Discussion}\label{discussion.sec}

\subsection{Dependence on Reconnection Height}

\begin{figure*}[htp]
    \centering
    \includegraphics[width=0.5\linewidth]{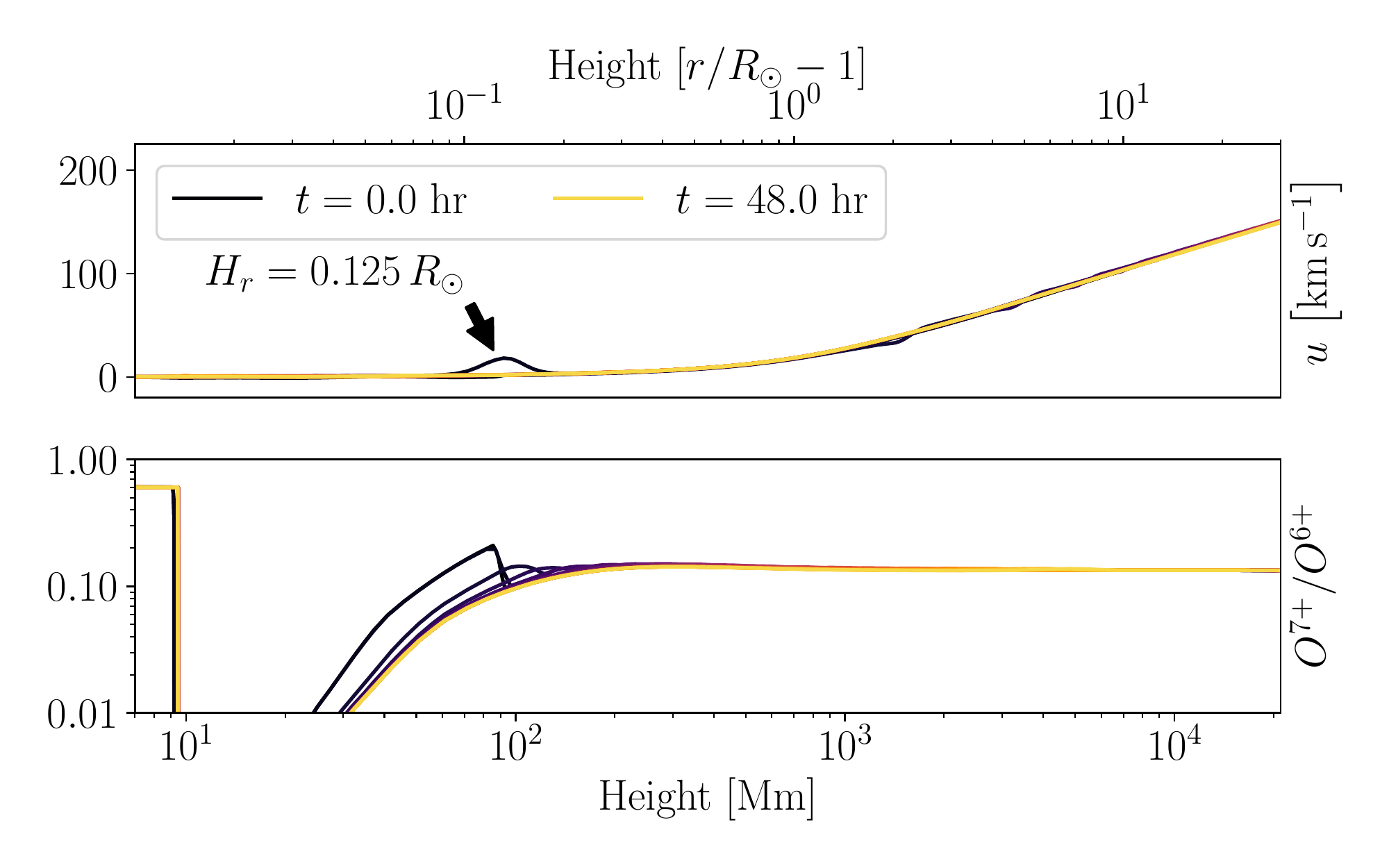}\includegraphics[width=0.5\linewidth]{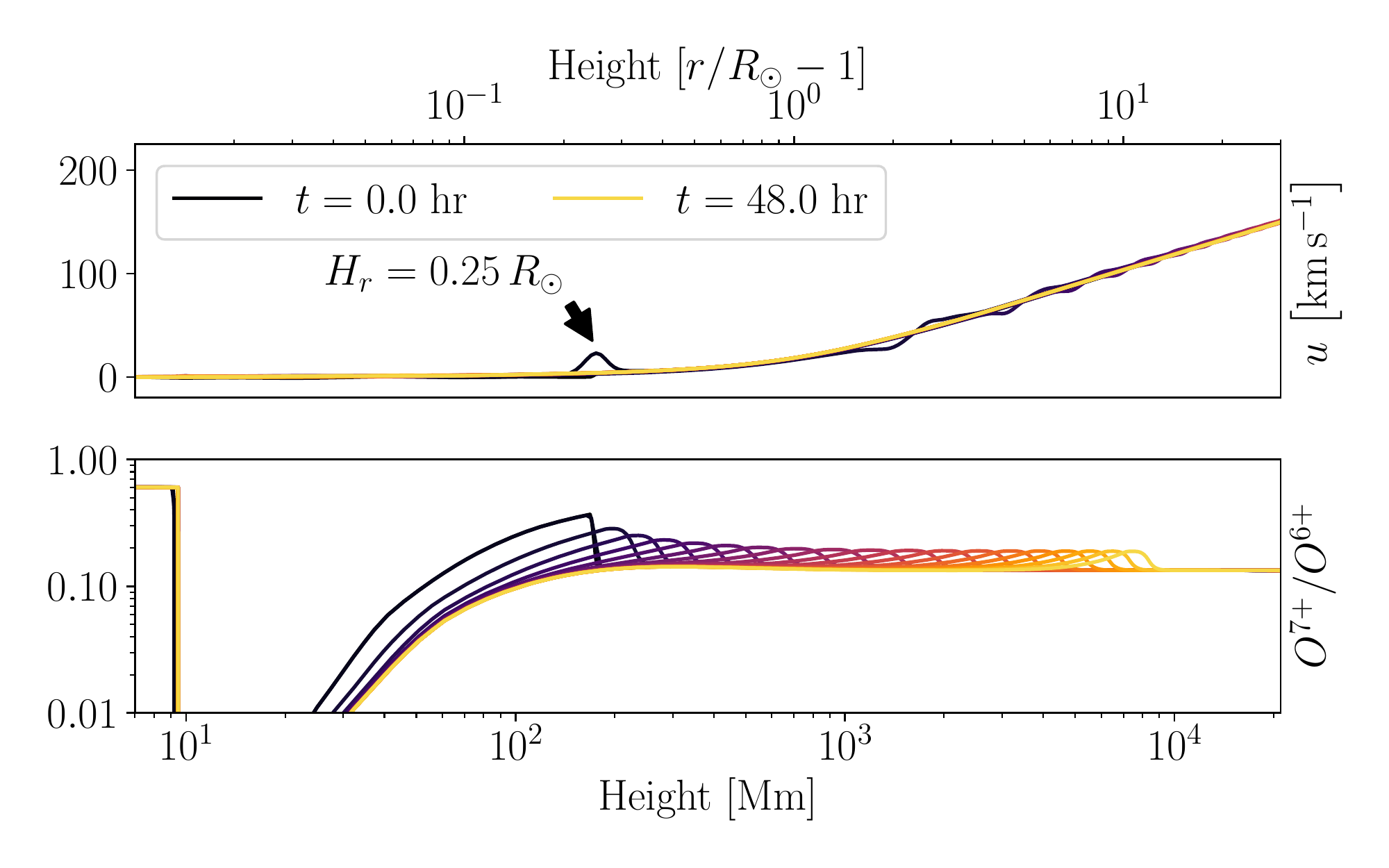}\\
    \includegraphics[width=0.5\linewidth]{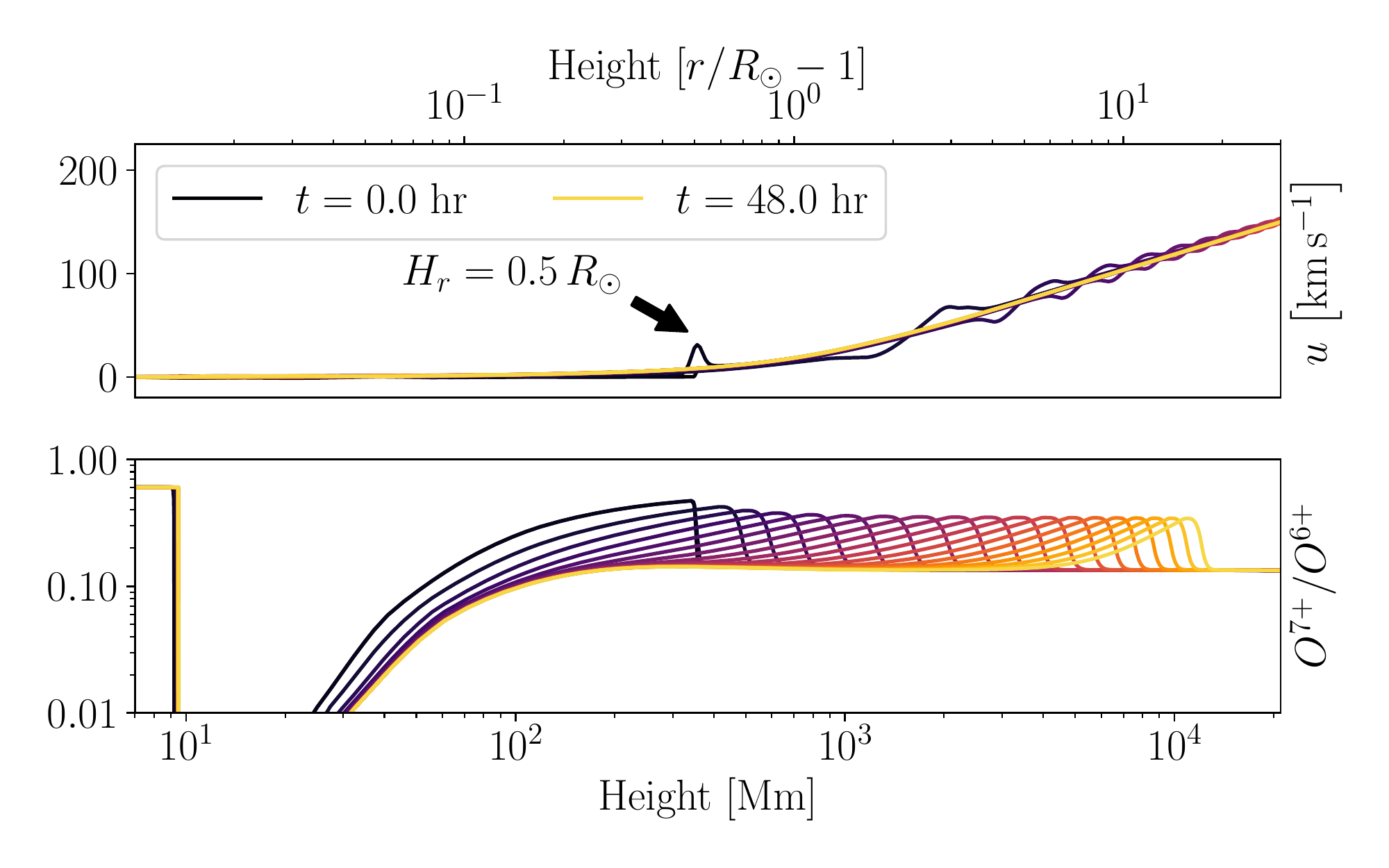}\includegraphics[width=0.5\linewidth]{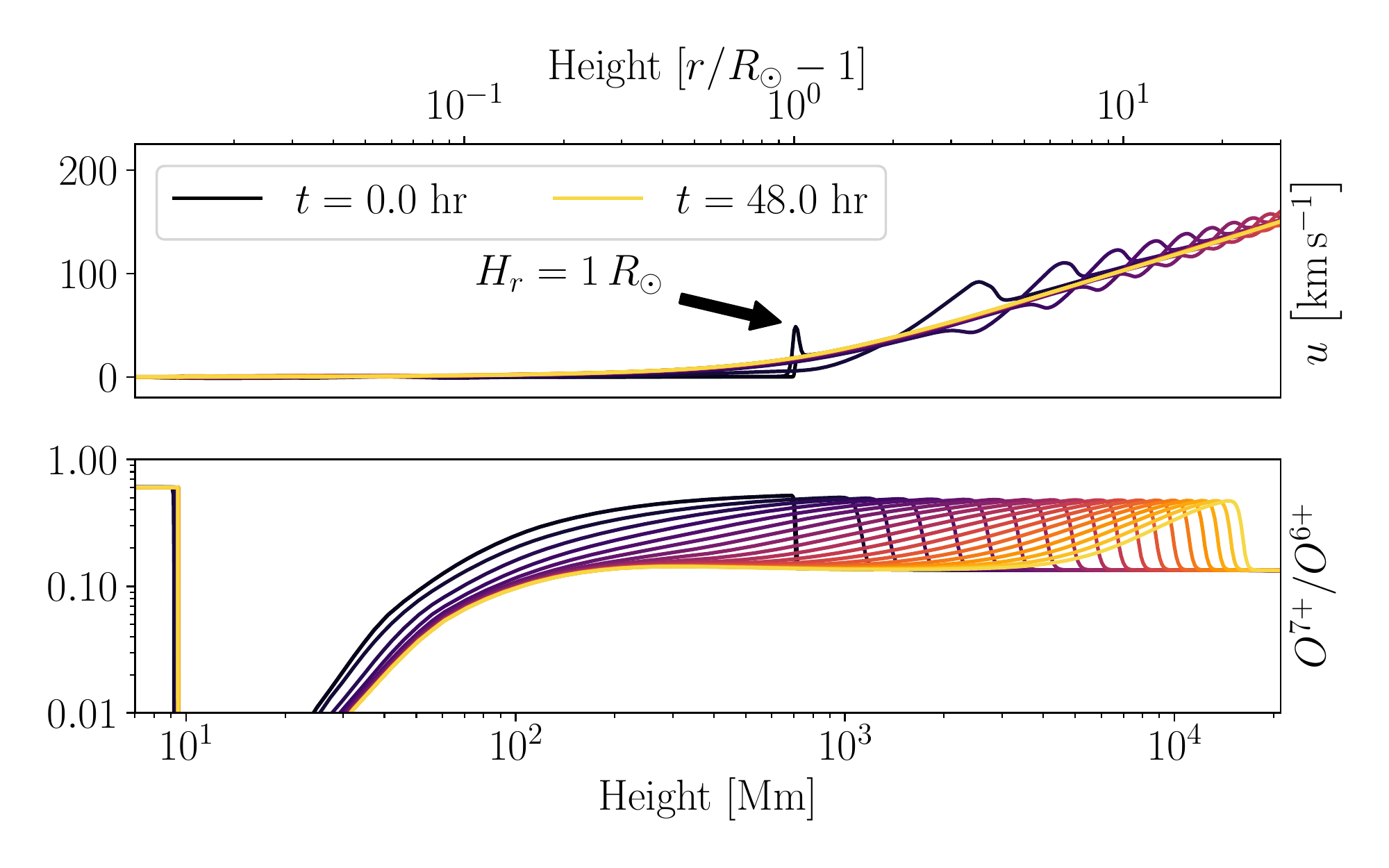}\\
    \includegraphics[width=0.5\linewidth]{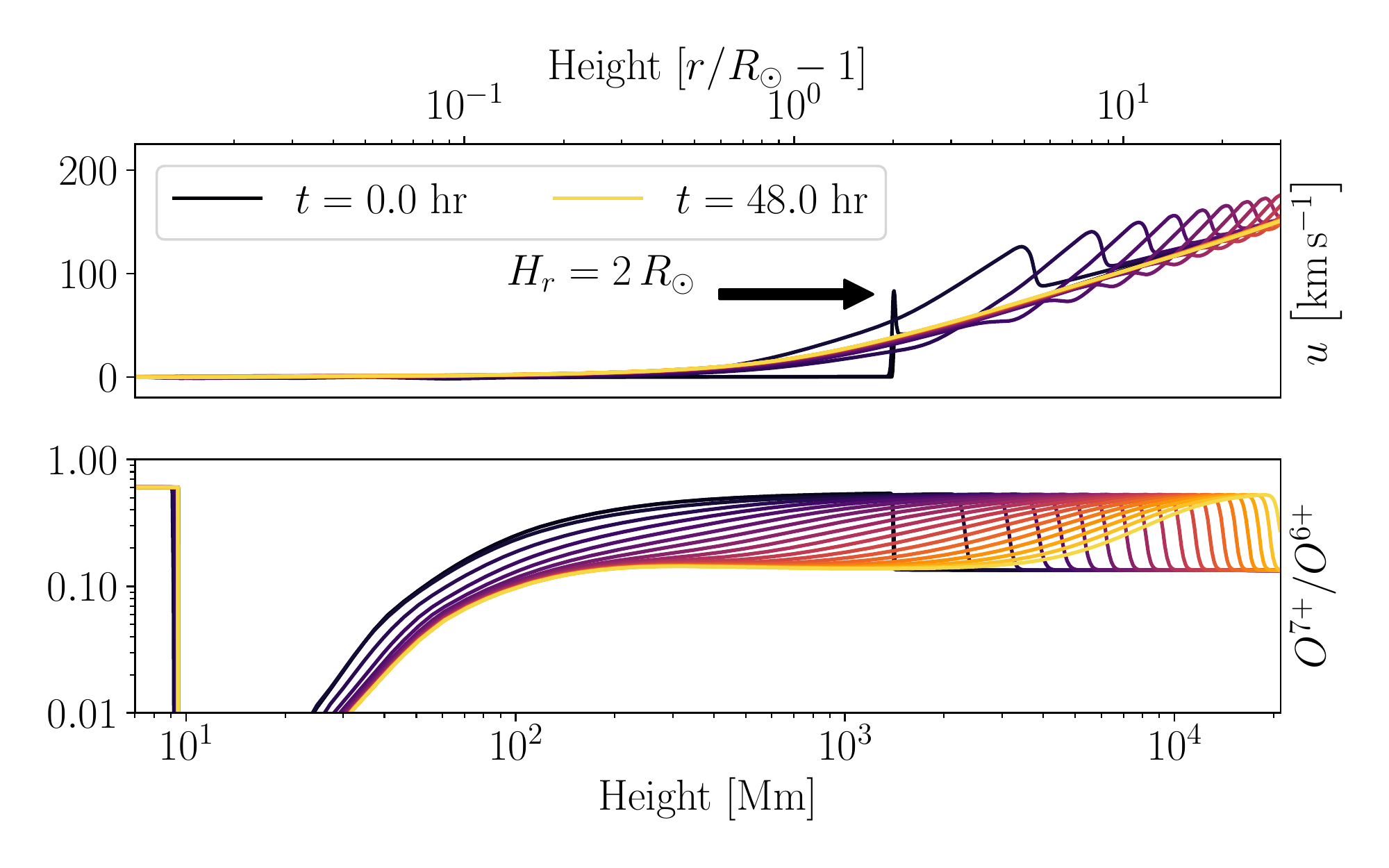}\includegraphics[width=0.5\linewidth]{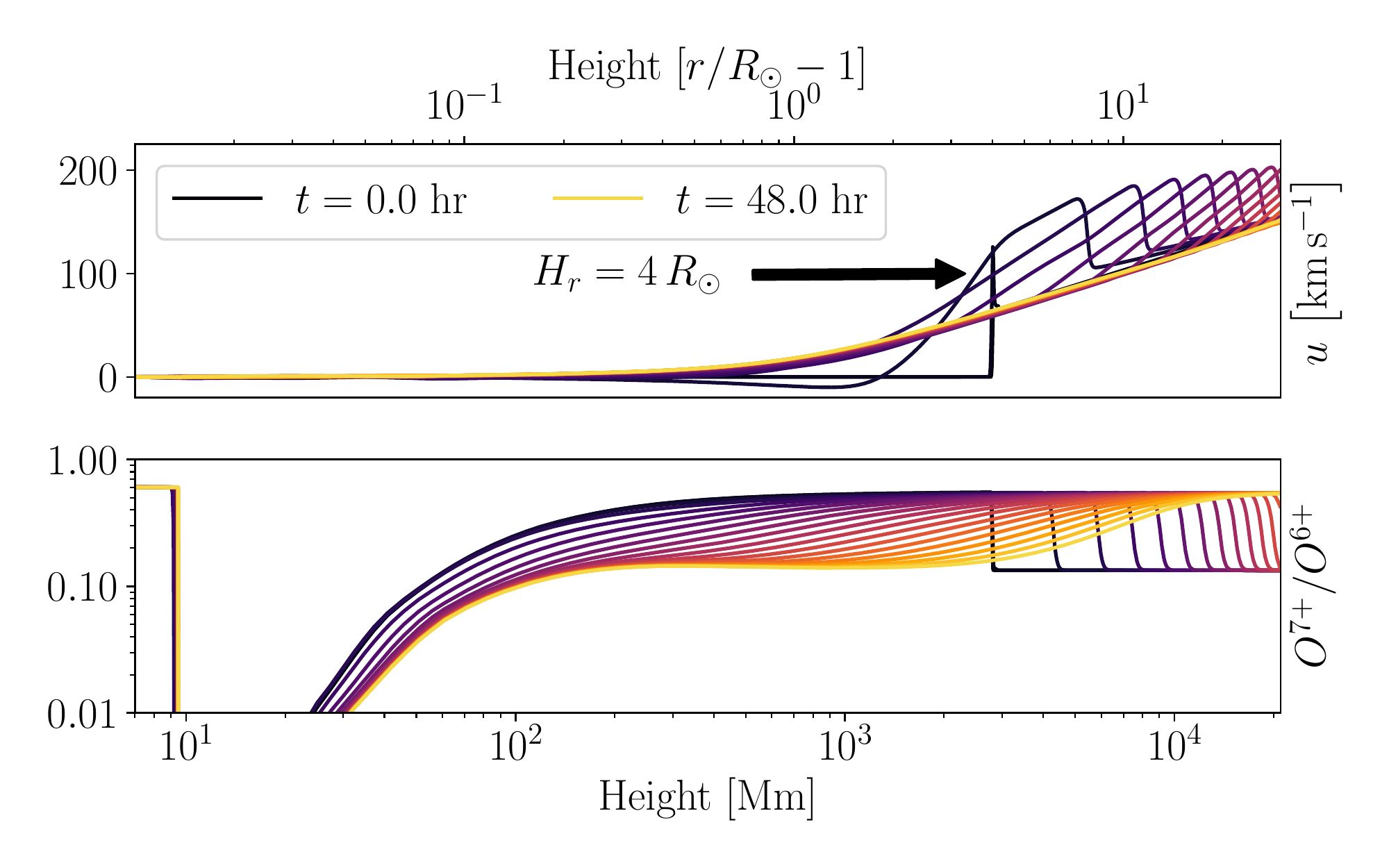}
    \caption{Comparison of flow speed and oxygen ionization ratio $\rm O^{7+}/O^{6+}$ for six separate experiments with reconnection sites placed at various heights from $R_\odot/8$ to $4R_\odot$. The dark purple to bright yellow colored curves represent snapshots at progressively later times from $0$ to $2\times10^5 {\rm s}$ (about $60$ hours) in increments of $10^4 {\, \rm s}$ (about $3$ hrs).}
    \label{fig.compare}
\end{figure*}

We explored the effects of reconnection height with six simulations that place the reconnection site in a variety of locations from $R_\odot / 8$ to $4 R_\odot$. 
A comparison of these runs is shown in Figure \ref{fig.compare}, which depicts the evolution of the fluid velocity and oxygen ionization ratio $\rm O^{7+}/O^{6+}$.
In each case, the strength and speed of the leading shock and trailing rarefaction depend primarily on the size of the initial discontinuity, which is larger for reconnection sites that are higher in the corona. 
Therefore, reconnection events that occur near or above a height of $R_\odot$ display an obvious velocity (and therefore density) signature, as this is the relevant height-scale for the Mach number of the wind solution to have appreciable size.
For reconnection sites that are well below this height ($H_r \le 0.5 R_\odot$) the initial discontinuity is too weak to create a strong shock and the fluid quickly settles onto a wind solution with little more than a short-lived transient wave disturbance.
The initial temperature change across the reconnection site is less strongly affected by changes in $H_r$; however, for reconnection sites lower in the corona the ion and electron temperatures are more collisionaly coupled, so the dissipation of the initial temperature jump by the electrons similarly smooths the ion temperature profile and the reconnection signature becomes very weak in the far-field. 

The ionization signature is similarly height dependent; however, in this case the relevant scale is whether the reconnection occurs above or below the freeze-in height, which occurs at roughly $R_\odot/3$. 
Reconnection events that occur above this height show a strong closed-field signature in the ionization ratios that is largely unaltered as it propagates into the heliosphere, followed by a slow decay back to the wind-like solution.
For reconnection sites that are lower in the corona, the ionization ratios undergo more evolution and the signature of the material from within the initially hydrostatic column becomes progressively weaker, so that for reconnection sites well below the freeze-in height there is little-to-no signature, with the plasma having undergone significant thermal-temporal evolution before reaching the heliosphere.
A critical difference, however, is that the ionization signature is not dissipated in the same way that the temperature and even velocity profiles can be, so whatever signature survives up to the freeze-in height is well preserved as it propagates into the heliosphere, as seen in the $H_r = 0.25\,R_\odot$ panel of Figure \ref{fig.compare}.
Note that while the relevant height for both fluid and ionization dynamics is of order $R_\odot$ in these simulations, these heights are model dependent, and may change for different energy deposition profiles and ion populations, such as Fe ($Z=26$), whose freeze-in height is likely to be higher than that of O ($Z=8$).

\subsection{Relevance to In Situ Observations}

\begin{figure*}[htp]
    \centering
    \includegraphics[width = 0.8\linewidth]{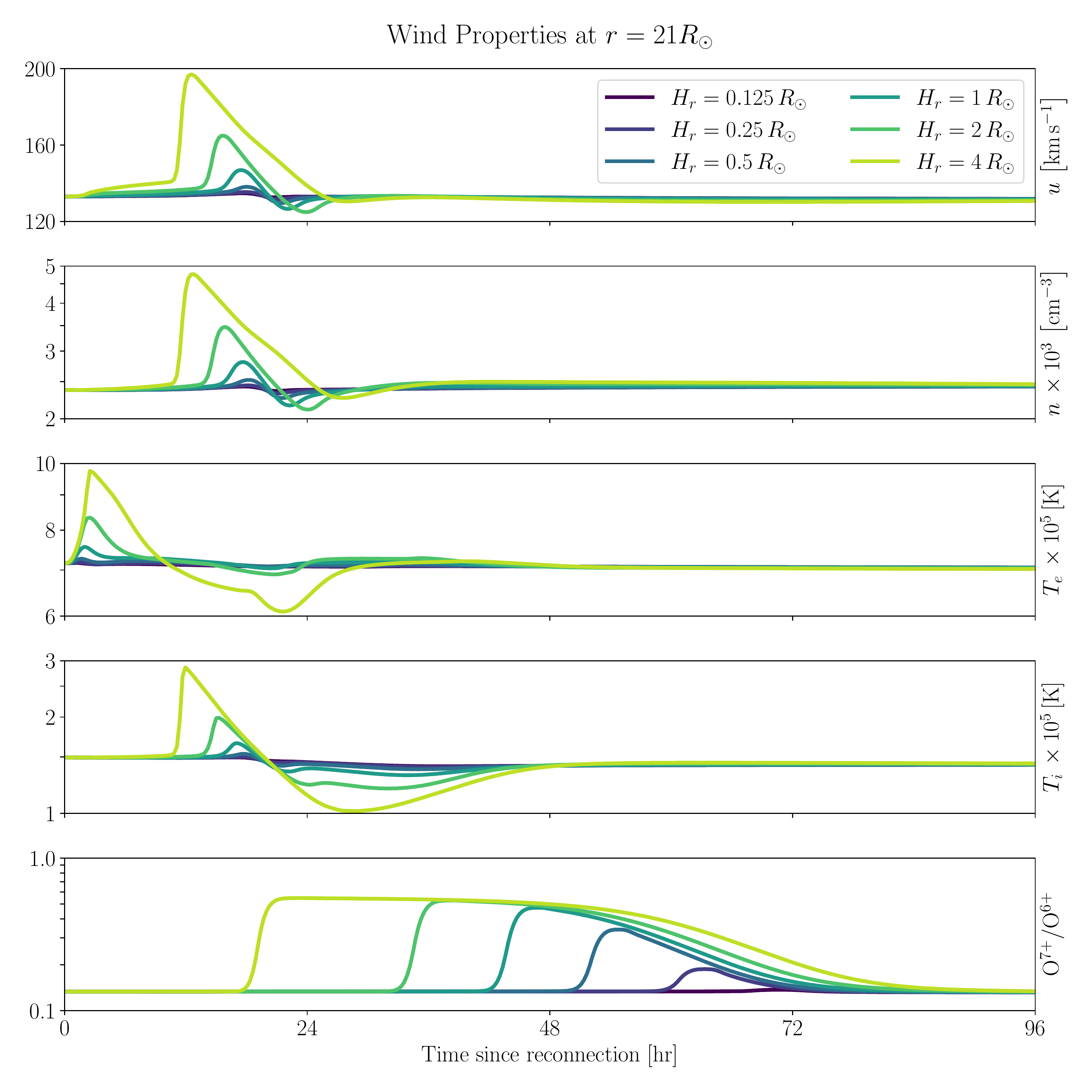}
    \caption{Time series of plasma properties at $r = 21 R_\odot$ for each of the runs. Note that in the time-series data the orientation of the ``N'' structures is reversed relative to the spatial profiles. The arrival time and duration of the N-waves reflect the acoustic travel time from the base of the corona, while the ionization ratio arrival time and duration reflects the fluid travel time and ionization/recombination time in the lower corona. For reconnection events below $\sim 0.25R_\odot$ there is little to no signature in the extended corona.}
    \label{fig.timing}
\end{figure*}

To explore the implications of these calculations for in situ observations we have extracted time series data for the fluid variables and ionization ratio $\rm O^{7+}/O^{6+}$ at $s=20\,R_\odot$ ($r = 21\, R_\odot$) above the solar surface, as shown in Figure \ref{fig.timing}.
This height is roughly coincident with the closest approach of Parker Solar Probe \cite[PSP,][]{Fox:2016d} during its sixth perihelion pass, and is at once far enough above the solar surface to be representative of the conditions in the heliosphere and also well within the numerical domain so as to avoid any possible boundary effects.
Unfortunately, PSP does not carry instrumentation to detect ionization ratios; however, these are not expected to undergo further evolution beyond $20\, R_\odot$, so the values and timing depicted in Figure \ref{fig.timing} should scale readily to heliocentric radii of $r > 60\, R_\odot$, at which point they will be detectable by Solar Oribiter \citep[SolO,][]{SolO}.

In the figure, results from the six simulations are indicated by the variously colored curves, which run from dark purple to bright yellow for reconnection sites of increasing height.
Consistent with the velocity profiles in Figure \ref{fig.compare}, the signatures in the density, temperature, and ionization ratios show consistently larger amplitudes as the reconnection height is increased.
The arrival time of these signatures is also earlier for reconnection events that are higher in the corona, due in part to the reduced travel distance from the reconnection site to $20\, R_\odot$.
For the most part, the fluid signatures arrive simultaneously, with the electron heat front being an exception due to the significantly larger electron thermal speed, which makes the transit time from the reconnection site to $20\, R_\odot$ difficult to resolve on the scales represented here.
For the remaining fluid properties, however, the arrival time is governed by the two remaining characteristic speeds -- the sound speed (ion acoustic speed) and the bulk flow speed, both of which are on the order of a few $10^2 \, \rm km\, s^{-1}$ over the majority of the spatial domain.

In the case of the density, velocity, and ion temperature, the primary signature is the leading shock, whose arrival time reflects the shock propagation speed from the reconnection site, through the lower and middle corona, and into the extended corona and heliosphere.
This speed depends on the strength of the shock, being equal to the sound speed for weak shocks and as much as a few times the sound speed for strong shocks.
Reconnection events that are closer to the transition region exhibit weaker/slower shocks that must propagate from lower in the corona, where the temperature and sound speed are correspondingly smaller.
For that reason the signatures from reconnection events higher in the corona are not only larger in amplitude, but also arrive earlier and last longer by as much as a several hours, with the $H_r = 4 R_\odot$ event spanning from $t\simeq12-30\,\rm hrs$ while the $H_r = R_\odot/8$ event occurs over a much shorter time from $t\simeq 20-22\,\rm hrs.$

Nonetheless, the qualitative signatures are identical in each case, with an initial enhancement across the leading shock, followed by a slow decay through the rarefaction and finally a small jump across the weak, trailing shock, which arrives later for the higher-amplitude events, owing to the larger extent of the rarefaction. 
This causes the signatures from reconnection events that are higher in the corona to be not only stronger but also longer-lived.
Critically, the transient flow speed within the N-wave is systematically higher than the ambient wind speed, so that the net effect of interchange reconnection is to increase the speed of the wind relative to a steady-state solution.

For the ionization ratios we see a similar trend, with reconnection events placed higher in the corona having a stronger and more long-lasting signature, that arrives sooner and takes longer to decay.
However, as we have previously shown the ionization ratios do not participate in the shock-compression dynamics, but are instead entrained behind the shock.
As a result the ionization signatures \rev{arrive systematically} later than the shock/rarefaction system, being typically delayed by several hours, enough that in most cases the fluid variables have returned to equilibrium values by the time the ionization signatures arrive.
Within the ionization curves we see qualitatively similar behavior in all cases, with an initial enhancement of $\rm O^{7+}/O^{6+}$ reflecting the population that originated from within the hydrostatic region above the freeze-in height but below the reconnection site, followed by a slow decay reflecting the populations that passed through the freeze-in region before the N-wave had fully formed, and finally a return to the wind conditions from populations that originated near the base of the corona and experienced a thermal history that reflects the fully-relaxed wind solution. 

From the study of \cite{Zhao:2017}, the ratio $\rm O^{7+}/O^{6+}$ in observed solar wind streams is typically in the range 0.05 to 0.2, being systematically lower for wind streams that are ballistically mapped from coronal hole regions vs active regions.
By comparison, the steady-state wind stream that we simulate here exhibits an $\rm O^{7+}/O^{6+}$ ratio of $\sim 0.11$, with enhancements following reconnection of up to $0.5$ in the most extreme case. 
This suggests that reconnection events that occur lower in the corona and below the freeze-in height may be better matched to the observations, as would be the case for the $H_r = 0.25\,R_\odot$ simulation, which exhibits a maximum $\rm O^{7+}/O^{6+}$ ratio of 0.2.

The relative timing of these signatures similarly reflects the amplitude of the N-wave, with larger shock rarefaction systems having faster flows in the rarefaction region, corresponding to earlier arrivals of the associated ionization profiles.
The location of the reconnection sites relative to the freeze-in height is further evidenced in the flatness of the ionization curves between the arrival of the reconnection signature and the slow return to the wind-like signature, as this reflects the extent of the material in the column that originated below $H_r$ but above the freeze-in height.
By comparison, for $H_r < R_\odot$ the initially hydrostatic populations have clearly undergone significant ionization and recombination on their way into the heliosphere, with the signatures of their origin in the closed-field domains being progressively weaker for lower reconnection heights.

\subsection{Interpretation in 2+ Dimensions}

\begin{figure}[ht]
    \centering
    \includegraphics[width=\linewidth]{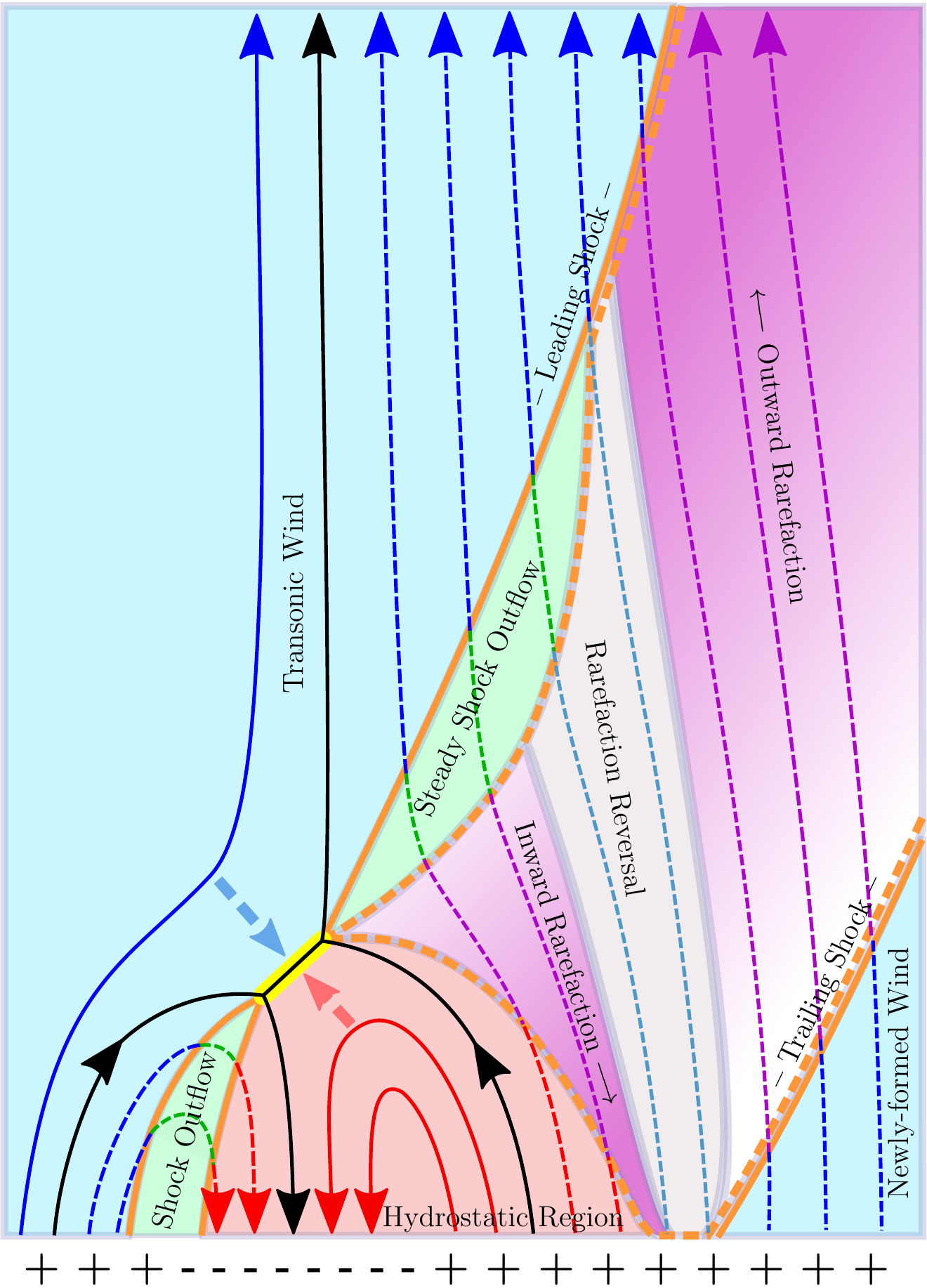}
    \caption{Schematic of 2D interchange reconnection with overlaid hydrodynamic sub-domains. Solid orange curves represent shock fronts while dashed orange curves are weak discontinuities at the leading and trailing edges of the rarefaction wave. In the upper-right outflow region, near the reconnection site, the leading shock separates the pre-shock wind from the post-shock outflow above the inward rarefaction, which propagates down into the hydrostatic column. Farther from the reconnection site the rarefaction undergoes reversal, eventually subsuming the post-shock outflow. A second shock separates the now-outward rarefaction from the newly-formed wind below it, and together with the leading shock these three features make up the N-wave.}
    \label{fig.2d}
\end{figure}

To this point we have focused on the time-dependent evolution the plasma along individual field lines;
however, if we return to the steady-state picture of interchange reconnection as depicted in Figure \ref{fig.rxn} we can imagine that all of the field lines in the outflow region(s) are undergoing the same dynamic evolution.
By associating the field-perpendicular distance from the reconnection site with increasing time since reconnection we can map the various stages of the field-aligned dynamics along a given field line to different subdomains within the reconnection outflow region.
A depiction of this mapping is shown in Figure \ref{fig.2d}, with the various hydrodynamic regions overlaid onto the magnetic field structure, which has been modified slightly from Figure \ref{fig.rxn} to reflect the collapse of the null-point into a thin current sheet.

Just as in Figure \ref{fig.rxn}, there are four magnetic domains, two inflow regions (I and IV) and two outflow regions (II and III). 
The pre-reconnection transonic and hydrostatic fluid domains encompass the two magnetic inflow regions, but also extend into the outflow regions.
This follows directly from the details of the field-aligned evolution and also reflects a broader causal connection between magnetic topology and plasma dynamics. 
Since field lines depend on the global structure of the magnetic field at a single instant in time, changes in magnetic connectivity occur instantaneously;
however, plasma dynamics are dictated by local processes and signals that propagate at finite speed, so the effects of reconnection are limited to the causally connected region defined by the characteristics of the system (i.e., the leading shock and inward rarefaction).

In the closed-field outflow region (domain III from Figure \ref{fig.rxn}) the reconnection discontinuity connects a segment of the wind solution from the lower corona (domain I) with a hydrostatic plasma column from the closed-field region beneath the null point (domain IV).
The resulting dynamics have not yet been simulated within our model; however, because the fluid velocity of the wind solution is directed toward the discontinuity, the resulting evolution will likely exhibit a pair of shocks that propagate away from each other and downward toward the solar surface.
These would then rebound from the density gradient at the base of the transition region, driving chromospheric evaporation, siphon flows, and other interesting processes in the lower corona.

In the open-field outflow region (domain II from figure \ref{fig.rxn}) the reconnection event connects an upper segment of the wind solution to a hydrostatic column below it and the dynamics along a given field line proceed as previously described. 
The various shocks and rarefactions along progressively older field lines form loci that generalize to magnetoacoustic disturbances. 
The leading shock and inward rarefaction along individual field lines generalize to slow magneto-acoustic fronts that separate the causally-disconnected portions of the outflow region in the far-field from the interior of the structure that develops from the initial reconnection discontinuity.
The reversal of the inward-propagating rarefaction, which subsequently subsumes the quasi-steady shock outflow, and the formation of the outward-propagating rarefaction and trailing shock each define subdomains as depicted by the variously colored regions in the figure. 
The final N-wave structure is defined by the two near-parallel shock fronts and the outward-propagating rarefaction that separates them, with the wind speed being again enhanced relative to the steady state within the interior of the N-wave.

It is noteworthy that in both of the outflow regions (the open-field outflow that we have modeled here and the closed-field outflow that we have not yet addressed) the effect of the fluid discontinuity should be to generate a pair of shock-like fronts that expand as a wedge away from the reconnection site.
This closely resembles the structure of a Petschek reconnection outflow, where a pair of slow magneto-acoustic shocks form in order to alter the effective aspect ratio of the current sheet to accommodate an increased reconnection rate over the classic Sweet-Parker scaling \citep{Parker:1957, Sweet:1958, Petschek:1964, Kulsrud:2001}.
Here the shocks result from an asymmetry in the conditions in the two inflow regions, including a non-zero velocity jump across the reconnection site; however, it seems likely that the two are related and may in fact be limiting cases of the same underlying phenomenon.
\rev{Other authors have previously observed the connection between one-dimensional Riemann problems and Petscheck reconnection, both symmetric and assymetric \citep[see][and references therin]{Lin:1999}, but these have generally focused on the cross-field dynamics in configurations where the asymptotic speed is zero in both inflow domains.
Clearly, there is more work to be done in unifying these descriptions under more general initial conditions as the resulting behavior will depend on the overall reconnection rate and the extent of the asymmetry.}

\section{Conclusions}\label{conclusions.sec}

We have demonstrated how a post-reconnection plasma discontinuity naturally evolves as a shock-rarefaction system in the manner of a Riemann problem, with a leading shock that expands into the heliosphere and an underlying rarefaction that propagates inward toward the solar surface.
We have further shown how the initially-downward-propagating rarefaction reverses at the transition region to become an outward-propagating shock-rarefaction system, which eventually overtakes the leading shock to form a shock-rarefaction-shock triplet, or N-wave.
This structure propagates coherently into the pre-existing wind in the heliosphere at a few times the local sound speed (or slow magneto-acoustic speed) after which the plasma behind it settles to a new quasi-steady wind solution on a time-scale that is comparable to the acoustic transit time from the solar surface to a given heliocentric radius.

\rev{A key finding of our model is that the release of material from the closed corona into the open field necessarily enhances the speed of the reconnected wind stream due to the interaction between the under-dense wind above the reconnection site and the high-pressure column below it.
Because the fast solar wind is generally steady while the slow wind exhibits significant variability, our interpretation is that interchange reconnection is more common along field lines that support slow wind streams than those that support fast wind streams.
This is consistent with the finding that the fast wind emanates from the interior of coronal holes while the slow wind originates from near coronal hole boundaries, where interchange reconnection is expected to be prevalent.
However, if our model is accurate then the specific processes that result in the slower baseline speed of the wind that emanates from coronal hole boundaries are independent of the process of interchange reconnection, whose primary effect is the create intermittent enhancements of the slow wind speed relative to the steady-state.}

Additionally, we have shown how the time-dependent ionization of oxygen evolves in the context of the post-reconnection fluid, with the ionization ratios being carried along by the flow while also evolving subject to the rate equations in the co-moving fluid frame.
We find that the location of the reconnection site relative to the freeze-in height is critical to the predicted temporal evolution of the ionization ratios for in situ measurements, with reconnection above the freeze-in height showing a strong signature of the initially hydrostatic plasma column that originates above the freeze-in height but below the reconnection site.
As the reconnection site is moved progressively closer to or below the freeze-in height, the initially hydrostatic fluid undergoes more ionization and recombination in its journey to the heliosphere, and so this signature is weakened considerably.

These observations have strong implications for the strength of the signatures from reconnection events that occur near large magnetic structures, whose vertical extent is comparable to the source surface radius (i.e., $H_r \sim 3 R_\odot$) vs. more compact structures with $H_r \lesssim R_\odot$.
Since pseudostreamers are usually on the lower end of this scale, it is likely that the strength of the discontinuity will be similarly on the lower end of the parameter space that we have described; however, because smaller structures are more likely to support the near-hydrostatic configuration on which our model depends, it is possible that the dependence on reconnection height will be weaker than these results suggest.

Despite this reduction in the strength of the ionization signal with decreased reconnection height, the time-delay between the arrival of the hydrodynamic and ionization signatures, which corresponds to the different travel times of the propagating signals (shocks and other waves) and the advected signals (material properties), seems to be robust across all reconnection heights.
And while we have not calculated them here, it is likely that disparities in abundances between open- and closed-field plasma will follow a similar trend, being similarly transported by the flow.
Confirmation of these features will depend on robust measurements of plasma dynamics, composition, and ionization in the inner heliosphere, and will require joint observation efforts from PSP and SolO, as the former will sample plasma in the appropriate source region, but only the latter possesses the suite of instruments required to make the measurement.
However, because the plasma's material composition is unlikely to be altered between $20 R_\odot$ and $200 R_\odot$ ($\sim 1 \rm AU$) it should be possible to reconstruct these signatures under ideal circumstances.

The applicability of these results depends on the existence of pristine initial conditions both above and below the reconnection site and minimal cross-field dynamics during the subsequent evolution.
The latter requirement will depend on the global magnetic evolution, which can be highly variable across disparate regions of the solar corona. 
However, these conditions are not unreasonable provided that the field-line drift velocity far from the reconnection site is not significantly greater than the acoustic speed and that the Lorentz force is sufficient to resist perpendicular gradients in the energy density of the fluid. 
That is to say that the coronal geometry should be relatively static and the plasma $\beta$ and Alfv\'en Mach numbers should both be small.

The validity of the initial condition depends additionally on the structure and dynamics of the reconnection site, \rev{which must be compact and and well defined with adjacent open- and closed-field domains being unambiguously identifiable.}
In particular, it is important that the thickness of the reconnection layer should not be significantly larger than the ion mean-free-path and that the fluid undergo minimal evolution during the time that it takes for a newly-reconnected field line to emerge from the reconnection site.
This will ensure that the fluid conditions on the newly reconnected field line appear as a discontinuity, which can then evolve in the manner that we have described.
\rev{Accordingly, while the slow solar wind is expected to emanate from both pseudostreamers and helmets streamers, the ambiguous nature of the open-closed boundary along helmet streamers and the continuous opening and closing of loops within the heliospheric current sheet may preclude the application of this model in that context; although a similar construction could be applicable given suitable alterations to the model.}

Assuming that the reconnection outflow speed is equal to the Alfv\'en speed, we can estimate the Alfv\'en transit time over the length of the reconnection site and compare it to the acoustic transit time over an ion mean-free-path length.
Combined with our earlier requirement on the thickness of the reconnection layer we then find that the aspect ratio of the reconnection site should be less than $\sqrt{2/\beta}$, where $\beta = 2 c_s^2 / v_a^2$ reflects the ratio of the acoustic speed to the Alfv\'en speed.
These assumptions are not universally applicable throughout the solar corona; however, they are sufficiently general as to be applicable in certain cases, and we have described a 2+D geometry that represents the likeliest scenario for observing this phenomenon, that being the case of laminar/coherent reconnection across a null-separatrix system that is representative of a simplified coronal pseudostreamer.
In this construction the expanding N-wave forms a wedge of high-speed wind within the rarefaction that is bounded by slow magneto-acoustic shocks at its leading and trailing edges.
Depending on the structure of the magnetic field ahead of the N-wave this may provide a framework for the formation of magnetic switchbacks vis-\`a-vis the model of \cite{Schwadron:2021m}.
Moreover, the structure of the reconnection outflow in this model is strongly reminiscent of Petscheck reconnection, with the acoustic shocks and rarefaction described here being naturally generalized to slow magnetosonic shocks and rarefactions.

In reality, the solar corona is always evolving, and that evolution will invariably complicate the behavior that we have demonstrated.
The sensitivity of these results to the assumptions within our model remain to be tested; however, the dynamics that we have described here should persist in some form anywhere that the basic magnetic geometry is comparable to our model framework and the reconnection process is not too violent.
As such, these dynamics should be viewed as a baseline for comparison so that the disparate effects of coherent reconnection and more complicated dynamic processes can be disentangled when interpreting in situ and remote sensing observations.

\section{Acknowledgements}\label{acknowledgements.sec}
RBS, SJB and MGL were supported for this project by NASA HSR grant 80HQTR21T0106. RBS and MGL were also supported for this project by the Office of Naval Research and by NASA PSP/WISPR grant NNG11EK11I. We thank the anonymous referee for their helpful comments and careful reading of the manuscript. Thanks go also to Peter Wyper for helpful discussions on the conditions near the reconnection site.

\bibliographystyle{apj}
\bibliography{bibfile.bib}

\newcommand{\noop}[1]{}
\begin{thebibliography}{}
\expandafter\ifx\csname natexlab\endcsname\relax\def\natexlab#1{#1}\fi

\bibitem[{{Aslanyan} {et~al.}(2021){Aslanyan}, {Pontin}, {Wyper}, {Scott},
  {Antiochos}, \& {DeVore}}]{Aslanyan:2021a}
{Aslanyan}, V., {Pontin}, D.~I., {Wyper}, P.~F., {et~al.} 2021, \apj, 909, 10

\bibitem[{Bradshaw {et~al.}(2011)Bradshaw, Aulanier, \&
  Del~Zanna}]{Bradshaw:2011d}
Bradshaw, S.~J., Aulanier, G., \& Del~Zanna, G. 2011, \apj, 743, 66

\bibitem[{{Bradshaw} \& {Cargill}(2013)}]{Bradshaw:2013}
{Bradshaw}, S.~J., \& {Cargill}, P.~J. 2013, \apj, 770, 12

\bibitem[{{Bradshaw} \& {Mason}(2003)}]{Bradshaw:2003a}
{Bradshaw}, S.~J., \& {Mason}, H.~E. 2003, \aap, 401, 699

\bibitem[{{Cranmer} {et~al.}(2007){Cranmer}, {van Ballegooijen}, \&
  {Edgar}}]{Cranmer:2007}
{Cranmer}, S.~R., {van Ballegooijen}, A.~A., \& {Edgar}, R.~J. 2007, \apjs,
  171, 520

\bibitem[{{Crooker} {et~al.}(2002){Crooker}, {Gosling}, \&
  {Kahler}}]{Crooker:2002}
{Crooker}, N.~U., {Gosling}, J.~T., \& {Kahler}, S.~W. 2002, \jgr, 107, 1028

\bibitem[{{Del Zanna} {et~al.}(2015){Del Zanna}, {Dere}, {Young}, {Landi}, \&
  {Mason}}]{DelZanna:2015}
{Del Zanna}, G., {Dere}, K.~P., {Young}, P.~R., {Landi}, E., \& {Mason}, H.~E.
  2015, \aap, 582, A56

\bibitem[{{Dere} {et~al.}(1997){Dere}, {Landi}, {Mason}, {Monsignori Fossi}, \&
  {Young}}]{Dere:1997}
{Dere}, K.~P., {Landi}, E., {Mason}, H.~E., {Monsignori Fossi}, B.~C., \&
  {Young}, P.~R. 1997, \aaps, 125, 149

\bibitem[{{Endeve} \& {Leer}(2001)}]{Endeve:2001m}
{Endeve}, E., \& {Leer}, E. 2001, \solphys, 200, 235

\bibitem[{{Fitzpatrick}(2015)}]{Fitzpatrick:2015}
{Fitzpatrick}, R. 2015, Plasma Physics: An Introduction (Florida: CRC Press)

\bibitem[{{Fox} {et~al.}(2016){Fox}, {Velli}, {Bale}, {Decker}, {Driesman},
  {Howard}, {Kasper}, {Kinnison}, {Kusterer}, {Lario}, {Lockwood}, {McComas},
  {Raouafi}, \& {Szabo}}]{Fox:2016d}
{Fox}, N.~J., {Velli}, M.~C., {Bale}, S.~D., {et~al.} 2016, \ssr, 204, 7

\bibitem[{Friedrichs(1948)}]{Friedrichs:1948}
Friedrichs, K.~O. 1948, Communications on Pure and Applied Mathematics, 1, 211

\bibitem[{{Gilly} \& {Cranmer}(2020)}]{Gilly:2020o}
{Gilly}, C.~R., \& {Cranmer}, S.~R. 2020, \apj, 901, 150

\bibitem[{{Grappin} {et~al.}(2011){Grappin}, {Wang}, \&
  {Pantellini}}]{Grappin:2011}
{Grappin}, R., {Wang}, Y.~M., \& {Pantellini}, F. 2011, \apj, 727, 30

\bibitem[{Higginson {et~al.}(2017)Higginson, Antiochos, DeVore, Wyper, \&
  Zurbuchen}]{Higginson:2017a}
Higginson, A.~K., Antiochos, S.~K., DeVore, C.~R., Wyper, P.~F., \& Zurbuchen,
  T.~H. 2017, \apj, 837, 113

\bibitem[{Huba(1998)}]{NRL_Formulary}
Huba, J. D. (Joseph D.~), . 1998, NRL Plasma Formulary (Revised 1998.
  Washington, DC : Naval Research Laboratory, {1998})

\bibitem[{{Kulsrud}(2001)}]{Kulsrud:2001}
{Kulsrud}, R.~M. 2001, Earth, Planets and Space, 53, 417

\bibitem[{{Laming} {et~al.}(2019){Laming}, {Vourlidas}, {Korendyke}, {Chua},
  {Cranmer}, {Ko}, {Kuroda}, {Provornikova}, {Raymond}, {Raouafi}, {Strachan},
  {Tun-Beltran}, {Weberg}, \& {Wood}}]{Laming:2019}
{Laming}, J.~M., {Vourlidas}, A., {Korendyke}, C., {et~al.} 2019, \apj, 879,
  124

\bibitem[{{Landi} {et~al.}(2012){Landi}, {Alexander}, {Gruesbeck}, {Gilbert},
  {Lepri}, {Manchester}, \& {Zurbuchen}}]{Landi:2012j}
{Landi}, E., {Alexander}, R.~L., {Gruesbeck}, J.~R., {et~al.} 2012, \apj, 744,
  100

\bibitem[{{Leer} {et~al.}(1982){Leer}, {Holzer}, \& {Fla}}]{Holzer:1982}
{Leer}, E., {Holzer}, T.~E., \& {Fla}, T. 1982, \ssr, 33, 161

\bibitem[{{Lin} \& {Lee}(1999)}]{Lin:1999}
{Lin}, Y., \& {Lee}, L.~C. 1999, Physics of Plasmas, 6, 3131

\bibitem[{{Longcope} \& {Bradshaw}(2010)}]{Longcope:2010a}
{Longcope}, D.~W., \& {Bradshaw}, S.~J. 2010, \apj, 718, 1491

\bibitem[{{Masson} {et~al.}(2014){Masson}, {McCauley}, {Golub}, {Reeves}, \&
  {DeLuca}}]{Masson:2014j}
{Masson}, S., {McCauley}, P., {Golub}, L., {Reeves}, K.~K., \& {DeLuca}, E.~E.
  2014, \apj, 787, 145

\bibitem[{{McComas} {et~al.}(2000){McComas}, {Barraclough}, {Funsten},
  {Gosling}, {Santiago-Mu{\~n}oz}, {Skoug}, {Goldstein}, {Neugebauer}, {Riley},
  \& {Balogh}}]{McComas:2000.105}
{McComas}, D.~J., {Barraclough}, B.~L., {Funsten}, H.~O., {et~al.} 2000, \jgr,
  105, 10419

\bibitem[{{M\"uller} {et~al.}(2020){M\"uller}, {St. Cyr, O. C.}, {Zouganelis,
  I.}, {Gilbert, H. R.}, {Marsden, R.}, {Nieves-Chinchilla, T.}, {Antonucci,
  E.}, {Auch\`ere, F.}, {Berghmans, D.}, {Horbury, T. S.}, {Howard, R. A.},
  {Krucker, S.}, {Maksimovic, M.}, {Owen, C. J.}, {Rochus, P.},
  {Rodriguez-Pacheco, J.}, {Romoli, M.}, {Solanki, S. K.}, {Bruno, R.},
  {Carlsson, M.}, {Fludra, A.}, {Harra, L.}, {Hassler, D. M.}, {Livi, S.},
  {Louarn, P.}, {Peter, H.}, {Sch\"uhle, U.}, {Teriaca, L.}, {del Toro Iniesta,
  J. C.}, {Wimmer-Schweingruber, R. F.}, {Marsch, E.}, {Velli, M.}, {De Groof,
  A.}, {Walsh, A.}, \& {Williams, D.}}]{SolO}
{M\"uller}, D., {St. Cyr, O. C.}, {Zouganelis, I.}, {et~al.} 2020, A\&A, 642,
  A1

\bibitem[{Nikoli{\'c}(2019)}]{Nikolic:2019}
Nikoli{\'c}, L. 2019, Space Weather, 17, 1293

\bibitem[{{Owocki} {et~al.}(1983){Owocki}, {Holzer}, \&
  {Hundhausen}}]{Owocki:1983}
{Owocki}, S.~P., {Holzer}, T.~E., \& {Hundhausen}, A.~J. 1983, \apj, 275, 354

\bibitem[{{Parker}(1957)}]{Parker:1957}
{Parker}, E.~N. 1957, \jgr, 62, 509

\bibitem[{{Petschek}(1964)}]{Petschek:1964}
{Petschek}, H.~E. 1964, NASA Special Publication, 50, 425

\bibitem[{Pinto {et~al.}(2009)Pinto, Grappin, Wang, \&
  L{\'e}orat}]{Pinto:2009a}
Pinto, R., Grappin, R., Wang, Y.~M., \& L{\'e}orat, J. 2009, \aap, 497, 537

\bibitem[{{Pontin} {et~al.}(2013){Pontin}, {Priest}, \&
  {Galsgaard}}]{Pontin:2013}
{Pontin}, D.~I., {Priest}, E.~R., \& {Galsgaard}, K. 2013, Astrophys.~J., 774,
  154

\bibitem[{{Priest} {et~al.}(2003){Priest}, {Hornig}, \&
  {Pontin}}]{Priest:2003ja}
{Priest}, E.~R., {Hornig}, G., \& {Pontin}, D.~I. 2003, \jgr (Space Physics),
  108, 1285

\bibitem[{{Schwadron} \& {McComas}(2021)}]{Schwadron:2021m}
{Schwadron}, N.~A., \& {McComas}, D.~J. 2021, \apj, 909, 95

\bibitem[{{Scott} {et~al.}(2021){Scott}, {Pontin}, {Antiochos}, {DeVore}, \&
  {Wyper}}]{Scott:2021a}
{Scott}, R.~B., {Pontin}, D.~I., {Antiochos}, S.~K., {DeVore}, C.~R., \&
  {Wyper}, P.~F. 2021, \apj, 913, 64

\bibitem[{{Sweet}(1958)}]{Sweet:1958}
{Sweet}, P.~A. 1958, in Electromagnetic Phenomena in Cosmical Physics, ed.
  B.~{Lehnert}, Vol.~6, 123

\bibitem[{{Zhao} {et~al.}(2017){Zhao}, {Landi}, {Lepri}, {Gilbert},
  {Zurbuchen}, {Fisk}, \& {Raines}}]{Zhao:2017}
{Zhao}, L., {Landi}, E., {Lepri}, S.~T., {et~al.} 2017, \apj, 846, 135

\end{thebibliography}

\appendix

\section{Rarefaction Waves in an Expanding Flux Tube}

Rarefaction waves in the solar corona propagate through a non-uniform medium that is structured by variations in temperature and density, and anisotropic due to the magnetic field and gravitational acceleration.
While an exact analytical treatment is not feasible, we can inform our understanding of their structure by considering the lowest order approximation given by a radially expanding, single-temperature isothermal fluid with sound speed $C_0$.
We assume that no natural length scale exists and so the evolution is self-similar, being described by a similarity variable $\varphi = z / C_0 t$, where $z$ is the spatial coordinate describing the interior of the wave.
The leading edge of the rarefaction is composed of a weak discontinuity, which propagates at the sound speed into the unperturbed medium in the rest frame of the fluid.
Then, by construction, the Mach number within the interior of the rarefaction is always given at the leading edge as $M(z=0)=1$, while the interior structure $M(z)$ is dictated by conservation of mass and momentum.

In the limit that the ambient flow speed and sound speed vary slowly compared to the interior structure of the rarefaction the spatial derivative obeys $\partial z / \partial s \sim 1$ and the continuity equation (expressed in terms of the linear mass density $\rho A$) is
\begin{equation}
    \partial_t (\rho A) + \partial_z (\rho u A) = 0,
\end{equation}
while the simplified momentum equation becomes
\begin{equation}
    \partial_t u + u \partial_z u + C_0^2 \partial_z \ln \rho + \partial_z U_g = 0,
\end{equation}
where $U_g$ is the gravitational potential. 
These equations can be transformed into their self-similar form with the substitutions
\begin{equation}
    \partial_s = (\varphi / z) \partial_\varphi
    \quad \text{and} \quad 
    \partial_t = -(\varphi / t)\partial_\varphi 
\end{equation}
after which the continuity and momentum equations are expressed directly in terms of $\varphi$ as
\begin{equation}
    \partial_\varphi \ln (\rho A) = - \frac{1}{M - \varphi} \partial_\varphi M
    \label{mass_ss.eq}
\end{equation}
and
\begin{equation}
    \left ( (M-\varphi) - \frac{1}{M-\varphi} \right ) \partial_\varphi M = \partial_\varphi \left ( \ln A - U_g/C_0^2 \right).
    \label{mom_ss.eq}
\end{equation}

The above expressions cannot be integrated directly; however, if we assume that the inhomogeneous terms on the right hand side of Eq. \eqref{mom_ss.eq} are small then the left hand side must vanish, which implies that 
\begin{equation}
    M \rightarrow \varphi \pm 1.
\end{equation}
Here we have chosen the positive root based on the physical requirement that the rarefaction should expand in time, rather than steepening as a shock does. Then, returning to the self-similar form of the continuity equation \eqref{mass_ss.eq}, we find that
\begin{equation}
    \rho A = (\rho A)_0 \exp(\mp \varphi).
\end{equation}

This result assumes that the length scale of the geometry is always longer than that of the Mach number, meaning that the wave must eventually propagate into a region of ever-increasing geometric scale before its internal size becomes too large.
This ensures that the change in internal pressure across the rarefaction is larger than the influence of gravity or expansion and the resulting dynamics mimic the homogeneous case.
In this limit the velocity increases linearly away from the leading edge of the wave while the linear mass density $(\rho A)$ decreases exponentially in the direction of increasing velocity.
In the context of the solar corona, this means that for radial rarefactions, whether inward- or outward-propagating, the \emph{linear}-mass-density ($\rho A$) decreases into the wave (away from the leading edge) irrespective of the direction of propagation, even as the \emph{volumetric}-mass-density ($\rho$) almost always decreases with height.

\end{document}